\RequirePackage{rotating}
\documentclass[11pt,reqno]{amsproc}
\allowdisplaybreaks
\usepackage[only,llbracket,rrbracket,llparenthesis,rrparenthesis]{stmaryrd} 
\usepackage{fullpage}
\usepackage[semicolon,square,authoryear]{natbib}
\numberwithin{equation}{section}
\usepackage{cite}
\usepackage{enumerate}
\usepackage{caption}
\usepackage{morefloats}

\usepackage{psfrag}
\usepackage{graphicx}
\DeclareGraphicsExtensions{.eps}
\usepackage{epstopdf}
\usepackage{subfigure}
\usepackage{float}
\usepackage{soul}
\usepackage[utf8]{inputenc}
\usepackage[table]{xcolor}
\usepackage{booktabs}
\usepackage{multicol}
\usepackage{longtable}
\usepackage{tcolorbox}
\usepackage[debug=false, colorlinks=true, pdfstartview=FitV, linkcolor=blue, citecolor=blue, urlcolor=blue]{hyperref}
\usepackage{rotating}

\usepackage{morefloats}

\newlength{\drop}
\definecolor{amethyst}{rgb}{0.6, 0.4, 0.8}
\definecolor{burgundy}{rgb}{0.5, 0.0, 0.13}

\title{Global sensitivity analysis of frequency 
band gaps in one-dimensional phononic 
crystals}

\author{\textbf{W.~Witarto}, \textbf{K.~B.~Nakshatrala}, and 
\textbf{Y.~L.~Mo} \\
{\small Department of Civil \& Environmental Engineering, University of Houston, Texas 77204.} 
\textbf{Correspondence to:}~knakshatrala@uh.edu
}

\begin{document}

\begin{titlepage}
  \drop=0.1\textheight
  \centering
  \vspace*{\baselineskip}
  \rule{\textwidth}{1.6pt}\vspace*{-\baselineskip}\vspace*{2pt}
  \rule{\textwidth}{0.2pt}\\[\baselineskip]
  {\LARGE \textbf{\color{burgundy}
  Global sensitivity analysis of frequency 
  band gaps \\[0.3\baselineskip]
  in one-dimensional phononic crystals}}\\[0.2\baselineskip]
    \rule{\textwidth}{0.4pt}\vspace*{-\baselineskip}\vspace{3.2pt}
    \rule{\textwidth}{1.6pt}\\[\baselineskip]
    \scshape
    An e-print of this paper is available on arXiv:~1807.06454. \par
    \vspace*{\baselineskip}
    Authored by \\[\baselineskip]
     {\Large W.~Witarto \par}
    {\itshape Graduate Student, Department of Civil and Environmental Engineering\\
    University of Houston, Houston, Texas 77204-4003. \par}
    \vspace*{\baselineskip}
    {\Large Kalyana~B.~Nakshatrala\par}
    {\itshape Department of Civil \& Environmental Engineering \\
    University of Houston, Houston, Texas 77204--4003. \\ 
    \textbf{phone:} +1-713-743-4418, \textbf{e-mail:} knakshatrala@uh.edu \\
    \textbf{website:} \url{http://www.cive.uh.edu/faculty/nakshatrala}\par}
    \vspace*{\baselineskip}
    {\Large Yi-Lung Mo \par}
    {\itshape Professor, Department of Civil and Environmental Engineering\\
    University of Houston, Houston, Texas 77204-4003. \\
    \textbf{phone:} +1-713-743-4274, \textbf{e-mail:} yilungmo@egr.uh.edu
    \par}
    \vfill
        {\scshape 2018} \\
        {\small Computational \& Applied Mechanics Laboratory} \par
\end{titlepage}

\begin{abstract}
Phononic crystals have been widely employed in many 
engineering fields, which is due to their unique feature 
of frequency band gaps. 
For example, their capability to filter out the incoming 
elastic waves, which include seismic waves, will have 
a significant impact on the seismic safety of nuclear 
infrastructure. 
In order to accurately design the desired frequency band gaps, one must pay attention on how the input parameters and the interaction of the parameters can affect the frequency band gaps. Global sensitivity analysis can decompose the dispersion relationship of the phononic crystals and screen the variance attributed to each of the parameters and the interaction between them. 
Prior to the application in one-dimensional (1D) phononic crystals, this paper will first review the theory of global sensitivity analysis using variance decomposition (Sobol’ sensitivity analysis). Afterwards, the sensitivity analysis is applied to study a simple mathematical model with three input variables for better understanding of the concept. Then, the sensitivity analysis is utilized to study the characteristic of the first frequency band gap in 1D phononic crystals with respect to the input parameters. This study reveals the quantified influence of the parameters and their correlation in determining the first frequency band gap. In addition, simple straight-forward design equations based on reduced Sobol' functions are proposed to easily estimate the first frequency band gap. Finally, the error associated with the proposed design equations is also addressed.
\end{abstract}


\maketitle

\vspace{-0.3in}
\section*{Nomenclature -- Non-dimensional quantities} 
\label{Sec:S0_Notation}
\begin{longtable*}{|p{.3\textwidth} | p{.5\textwidth}|} \hline
  \small \\[-12pt]
  $\hat{u}_n=u_n/L_*=u_n/h$ & Displacement at layer $n$\\[-14pt] \\ \hline
  \\[-12pt]
  $\hat{h}_n=h_n/L_*=h_n/h$ & Height or thickness of layer $n$\\[-14pt] \\ \hline
  \\[-12pt]
  $\hat{h}=h/L_*=1$ & Total height or thickness of unit cell\\[-14pt] \\ \hline
  \\[-12pt]
  $\hat{z}_n=z/L_*=z_n/h$ &
  Position of interest within the layer with \\
  & reference to the bottom of layer $n$\\[-14pt] \\ \hline
  \\[-12pt]
  $\hat{k}_n=k_nL_*=kh$ & Wave number\\[-14pt] \\ \hline
  \\[-12pt]
  $\hat{t}=\dfrac{t}{T_*}=\dfrac{t}{h}\sqrt{\dfrac{E_1}{\rho_1}}$
  & Time\\[-14pt] \\ \hline
  \\[-12pt]
  $\hat{\omega}=\omega T_*=\omega h\sqrt{\rho_1/E_1} $
  & Radial frequency\\[-14pt] \\ \hline
  \\[-12pt]
  $\hat{\rho}_n=\rho_nL_*^3/M_*=\rho_n/\rho_1 $
  & Density of layer $n$\\[-14pt] \\ \hline
  \\[-12pt]
  $\hat{\lambda}_n=\lambda_nL_*T_*^2/M_*=\lambda_n/E_1 $
  & First Lam{\'e} parameter constant of layer $n$\\[-14pt] \\ \hline
  \\[-12pt]
  $\hat{\mu}_n=\mu_nL_*T_*^2/M_*=\mu_n/E_1 $
  & Second Lam{\'e} parameter constant of layer $n$\\[-14pt] \\ \hline
  \\[-12pt]
  $\hat{\sigma}_n=\sigma_nL_*T_*^2/M_*=\sigma_n/E_1 $
  & Normal stress at layer $n$\\[-14pt] \\ \hline
  \\[-12pt]
  $\hat{\tau}_n=\tau_nL_*T_*^2/M_*=\tau_n/E_1$
  & Shear stress at layer $n$\\[-14pt] \\ \hline
\end{longtable*}

\section{INTRODUCTION AND MOTIVATION}
\label{Sec:S1_ICs_Intro}
The study of wave propagation in the solid state physics has developed the so-called phononic crystals, an artificially fabricated periodic elastic structures \citep{kittel1996introduction}. These crystals exhibit unique properties of frequency band gaps, where incoming waves with frequencies falling inside the frequency band gaps will be reflected \citep{sigalas1992elastic,kushwaha1993acoustic,sigalas2005classical}. Depending on the structural scale, phononic crystals can be utilized to isolate different types of waves. At nanometer scales, the phononic crystals can effectively control heat propagation as the heat vibration oscillates at frequencies of the order of terahertz \citep{yu2010reduction}. At centimeter to micrometer scales, the frequency band gaps of the phononic crystals can isolate acoustic waves that oscillate at relatively lower frequencies (kilohertz to megahertz) \citep{martinez1995sound,gorishnyy2005hypersonic,maldovan2013sound}. At decimeter scales and larger, phononic crystals have been applied to isolate seismic waves \citep{brule2014experiments,xiang2012periodic,yan2014seismic,yan2015three}, which have frequencies lower than 50 Hz. The phononic crystals, therefore, can be engineered with different sizes and material constituents to cover any desired frequency ranges for different purposes.

The above-mentioned benefits denote an immense potential engineering applications of phononic crystals. In one of the most recent applications, the phononic crystals were utilized to enhance the seismic safety of nuclear power plant structures. An experimental study shows that a structural foundation designed using the concept of phononic crystals can successfully filter out the damaging frequency of input seismic waves and protect a small modular reactor building \citep{witarto2018seismic}. The phononic crystals can also be used as a non-intrusive method to protect existing structures in a form of wave barriers \citep{liu2015comparison}, which is highly desirable for existing nuclear power plant structures. The wave barriers can be used to sustain the unprecedented earthquake events without changing or retrofitting the existing nuclear power plant structures.

One of the challenging issues in designing the phononic crystals is the large number of random parameters as the input variables, which makes the problem high-dimensional. The inherent uncertainties associated with the input parameters and the interaction between parameters can makes the design process challenging and time-consuming. The generic approach to such problem is to identify the most influential input parameters and focus on those parameters in the design process. Although, parametric studies based on one-at-a-time technique have been performed to investigate how each of the material and geometric properties of phononic crystals may affect the frequency band gaps \citep{cheng2013novel,witarto2016analysis,bao2012dynamic}, the studies do not quantitatively rank the influence of each parameter nor consider the interaction between the parameters.

Global sensitivity analysis or analysis of
variance (ANOVA) based on Sobol’ decomposition
\citep{sobol1993sensitivity,sobol2001global} can
quantify the amount of variance that each of the
parameters and the interaction of two or more
parameters contribute to the mathematical model
output. This aspect will be breifly described in 
\textbf{Sec.~\ref{Sec:S2_Sobol_Notation}}.
We employ the global sensitivity analysis using Sobol’
decomposition on one-dimensional (1D) phononic crystals
(\textbf{Sec.~\ref{Sec:S3_Sobol_on_phononic}}).
This sensitivity analysis study will focus on the affecting
parameters on the first frequency band gap (i.e., the
lower bound frequency or the starting of
the frequency band gap) and the width of
the frequency band gap,
subjected to each of transverse wave (S-Wave) and longitudinal wave (P-Wave) excitations. Based on the most influential parameters, reduced models were derived as the simplified design equations to easily estimate the first frequency band gap of each of S-Wave and P-Wave.
A set of design equations (i.e., reduced-order models)
are derived in dimensionless form so that they can be
applied for design of 1D phononic crystals
at any scale (\textbf{Sec.~\ref{Sec:S4_Sobol_phononic_design}}).
Finally, concluding remarks, which include a discussion
on the dominant medium properties affecting the frequency
band gaps, are presented in
\textbf{Sec.~\ref{Sec:S5_Sobol_CR}}.

 Throughout this paper, a repeated index
 does not imply a summation over the index
 (i.e., we do not employ the Einstein's
 summation convention).

\section{THE SOBOL’ SENSITIVITY ANALYSIS}
\label{Sec:S2_Sobol_Notation}
Sobol’ sensitivity analysis is a global sensitivity analysis
method using variance decomposition that can handle linear
and nonlinear mathematical models. To make this paper
self-contained, a brief description of the Sobol’
decomposition is presented. Additional information
of the theory and assumptions can be found in the
original papers \citep{sobol1993sensitivity,sobol2001global}.

Consider a mathematical model which is
abstractly represented by the following
function:
\begin{align}
  \label{Eqn:Sobol_mathematical_model}
  Y = F(\mathbf{x}) 
\end{align}
where $\mathbf{x} = (x_1,\cdots, x_n)$
is a set of input parameters on the
$n$-dimensional unit hypercube domain
\[
\Omega^{n} := \{\mathbf{x} \; \vert
\; 0 \leq x_i \leq 1, i = 1, \cdots, n\}.
\]
The mathematical model Eq.~\eqref{Eqn:Sobol_mathematical_model}
can be decomposed into a series of increasing order Sobol’
functions as follows:
\begin{align}
  \label{Eqn:Sobol_Sobol_decomposition}
  F(\mathbf{x}) = F_0 + \sum_{i=1}^{n} F_{i}(x_i)
  + \sum_{i=1}^{n} \sum_{j = i + 1}^{n} F_{ij}(x_i,x_j)
  + \cdots + F_{1\cdots n}(x_1,\cdots, x_n)
\end{align}
For Eq.~\eqref{Eqn:Sobol_Sobol_decomposition}
to hold, the following three criteria must be
satisfied:

\begin{enumerate}[(a)]
\item The first term of the right-hand side
  in Eq.~\eqref{Eqn:Sobol_Sobol_decomposition},
  $F_0$, must be a constant.
\item The integral of every summand over
  its own variables must be zero. That is,
  \begin{align}
    \int_{0}^{1} F_{i_1 \cdots i_s}(x_{i_1},
    \cdots, x_{i_s}) dx_{k} = 0
    \quad \forall k = i_1, \cdots, i_s
  \end{align}
\item The summands are orthogonal, which means that
  if $(i_1,\cdots,i_s) \neq (j_1,\cdots,j_t)$ then
  \begin{align}
    \int_{\Omega^{n}} F_{i_1\cdots i_s} F_{j_1\cdots j_t}
    d\mathbf{x} = 0 
  \end{align}
\end{enumerate}

Upon satisfying those three criteria, the individual member of the Sobol’ functions in Eq.~\eqref{Eqn:Sobol_Sobol_decomposition} can be calculated as follows
\begin{align}
\label{Eqn:Sobol_F_0}
    F_0 = \int_{\Omega^{n}}F(\mathbf{x}) d\mathbf{x}
\end{align}
\begin{align}
  \label{Eqn:Sobol_F_i}
  F_i(x_i) = \int_{\Omega^{n-1}}F(x_i,\mathbf{x}_{\sim i})
  d\mathbf{x}_{\sim i}-F_0
\end{align}
\begin{align}
  \label{Eqn:Sobol_F_ij}
  F_{ij}(x_i,x_j)=\int_{\Omega^{n-2}}F(x_i,x_j,\mathbf{x}_{\sim ij})
  d\mathbf{x}_{\sim ij}-F_i(x_i)-F_j(x_j)-F_0
\end{align}
where $\mathbf{x}_{\sim i}$ is the vector
corresponding to all variables except
$x_i$ in the input set $\mathbf{x}$.
Similarly $\mathbf{x}_{\sim ij}$ is the
vector corresponding to all variables
except $x_i$ and $x_j$ in the input set
$\mathbf{x}$. The higher order Sobol’
functions can be calculated in similar
manner.

The total variance of $F(\mathbf{x})$
can be defined as
\begin{align}
  \label{Eqn:Sobol_D}
  D = \int_{\Omega^{n}}
  F^2(\mathbf{x})d\mathbf{x}
  - {F_0}^2
\end{align}
which can be decomposed into partial variances as shown in Eq.~\eqref{Eqn:Sobol_D_decomposition}. The partial variances are associated with the Sobol’ functions and can be calculated by integrating the corresponding functions. The calculation of the first and second order variances are shown in Eqs. ~\eqref{Eqn:Sobol_D_i} and ~\eqref{Eqn:Sobol_D_ij}. The higher order variances can be calculated by integrating the higher order Sobol’ functions.
%
\begin{align}
\label{Eqn:Sobol_D_decomposition}
  D= \sum_{i=1}^{n} D_{i}
  + \sum_{i=1}^{n} \sum_{j = i + 1}^{n} D_{ij}
  + \cdots + D_{1\cdots n}
\end{align}
\begin{align}
\label{Eqn:Sobol_D_i}
  D_i=\int_{\Omega^{1}}F_i^2(x_i)dx_i
\end{align}
\begin{align}
\label{Eqn:Sobol_D_ij}
  D_{ij}=\int_{\Omega^{2}}F_{ij}^2(x_i,x_j)dx_idx_j
\end{align}

Using the individual partial variance, one can calculate the contribution of each variance to the total output. The contribution, known as the Sobol’ sensitivity indices, is characterized by the ratio of the partial variance relative to the total variance, as shown in Eqs. ~\eqref{Eqn:Sobol_S_i} and \eqref{Eqn:Sobol_S_ij}.
\begin{align}
\label{Eqn:Sobol_S_i}
 S_i= \dfrac{D_i}{D}
\end{align}
\begin{align}
\label{Eqn:Sobol_S_ij}
 S_{ij}= \dfrac{D_{ij}}{D}
\end{align}
The higher order Sobol’ indices can be calculated in similar approach. Therefore, the total Sobol’ indices is
\begin{align}
\label{Eqn:Sobol_sum_S}
   \sum_{i=1}^{n} S_{i}
  + \sum_{i=1}^{n} \sum_{j = i + 1}^{n} S_{ij}
  + \cdots + S_{1\cdots n}=1
\end{align}

In many cases where the model functions are complex and nonlinear, the analytical solutions may not available. Therefore, the integration can be approximated using Monte Carlo based numerical integration. In this numerical approach, the partial variances can be calculated without the need to evaluate Sobol’ functions beforehand \citep{sobol1993sensitivity}. For this direct estimation, two sets of input need to be generated; the original input set $\mathbf{x}$ and the complementary input set $\mathbf{x^c}$. The formulae for total, first order and second order variances using Monte Carlo estimation are
shown in Eqs.~\eqref{Eqn:Sobol_Monte_Carlo_D}--\eqref{Eqn:Sobol_Monte_Carlo_D_ij}.
\begin{align}
\label{Eqn:Sobol_Monte_Carlo_F_0}
\overline{F}_0 = \dfrac{1}{N}
\sum_{m=1}^{N} F(x_m)
\end{align}
%
\begin{align}
  \label{Eqn:Sobol_Monte_Carlo_D}
  \overline{D} = \dfrac{1}{N} \sum_{m=1}^{N}
  F^2(x_m) - {\overline{F}_0}^2
\end{align}
\begin{align}
  \label{Eqn:Sobol_Monte_Carlo_D_i}
  \overline{D}_i = \dfrac{1}{N}
  \sum_{m=1}^{N} F(x_m)F(x_{im},\mathbf{x}_{\sim im}^c)
  - {\overline{F}_0}^2
\end{align}
\begin{align}
  \label{Eqn:Sobol_Monte_Carlo_D_ij}
  \overline{D}_{ij} = \dfrac{1}{N}
  \sum_{m=1}^{N} F(x_m) F(x_{im},x_{jm},\mathbf{x}_{\sim ijm}^c)
  - \overline{D}_i - \overline{D}_j
  -{\overline{F}_0}^2
\end{align}
where $m$ represents ordinal number of a test and $N$ is the sample size of the Monte Carlo estimation. The $\mathbf{x}_{\sim i}^c$ is the vector corresponding to all variables except $x_i$ in the input set $\mathbf{x^c}$. Similarly, $\mathbf{x}_{\sim ij}^c$ is the vector corresponding to all variables except $x_i$ and $x_j$ in the input set $\mathbf{x^c}$. The bar on the symbol, such as $\overline{D}$, denotes that the expression is numerically integrated. The Sobol’ indices can then be evaluated with Eqs.~\eqref{Eqn:Sobol_S_i} and \eqref{Eqn:Sobol_S_ij} using the numerically estimated variances.

\subsection{An application on a simple mathematical model}
\label{Sec:S3_Sobol_on_simple_model}
In this section, we illustrate the Sobol'
sensitivity analysis using a simple
mathematical model. This model has been
adopted from \citep{arwade2010variance},
which also provides an analytical solution.
Herein, we adopt this model and examine
how the Sobol' functions and Sobol'
indices that are estimated using Monte
Carlo simulation compare with the
analytical solution. Such a study will
be particularly relevant to our paper,
as our model does not have an analytical
and we have to rely on Monte Carlo
simulations.

The simple mathematical model is
given by the following polynomial
representation:
\begin{align}
  \label{Eqn:Sobol_simple_model}
  F(x_1,x_2,x_3) = x_1^2 + x_2^4 + x_1 x_2
  + x_2 x_3^4
\end{align}
where $x_1$, $x_2$ and $x_3$ are
independent variables, each of
which is uniformly distributed
in $[-4,4]$ input space. To
perform the Sobol’ decomposition,
the input space of each of the
input variables needs to be
rescaled to $[-1,1]$. To this
end, the variables $x_1$, $x_2$
and $x_3$ are expressed as
$8y_1 - 4$, $8y_2 - 4$ and
$8y_3 - 4$, respectively,
where $y_1$, $y_2$ and $y_3$
are uniformly distributed in
$[0,1]$.

The analytical solution of the
Sobol' functions and the Sobol'
indices of the problem posed in
Eq.~\eqref{Eqn:Sobol_simple_model}
are provided in Table 
\ref{Table:Sobol_Analytical_Sobol_functions}
and Table \ref{Table:Sobol_Analytical_Sobol_indices}.

\begin{table}[H]
  \caption{Analytically derived Sobol'
    functions\label{Table:Sobol_Analytical_Sobol_functions}}
  \begin{tabular}{|c|c|} \hline
    Sobol' functions & Function expression \\ \hline
    $F_{0}$ & 56.533 \\
    $F_{1}$ & $x_1^2 - 5.333$ \\
    $F_{2}$ & $x_2^4 + 51.2 x_2 - 51.2$ \\
    $F_{3}$ & $0$ \\
    $F_{12}$ & $x_1 x_2$ \\
    $F_{13}$ & $0$ \\
    $F_{23}$ & $x_2 x_3^4 - 51.2 x_2$ \\
    $F_{123}$ & $0$ \\
    \hline 
  \end{tabular}
\end{table}

\begin{table}[H]
  \caption{Analytically calculated Sobol'
    indices\label{Table:Sobol_Analytical_Sobol_indices}}
  \begin{tabular}{|c|c|} \hline
    Sobol' indices & Index number \\ \hline
    $S_{1}$ & $0.0005$ \\
    $S_{2}$ & $0.4281$ \\
    $S_{3}$ & $0.0000$ \\
    $S_{12}$ & $0.0007$ \\
    $S_{13}$ & $0.0000$ \\
    $S_{23}$ & $0.5708$ \\
    $S_{123}$ & $0.0000$ \\
    \hline 
  \end{tabular}
\end{table}

It is observed that $S_2$ and $S_{23}$ are the dominant indices, which means that the individual variable $x_2$ and the interaction of variables $x_2$ and $x_3$ are the most influential to the model outcome. One can also find that there are no individual stands of $x_3$, combination $(x_1,x_3)$ and correlation between all three variables $(x_1,x_2,x_3)$ in the mathematical model. Therefore, it is expected to have their Sobol’ functions $F_3$, $F_{13}$ and $F_{123}$ and their corresponding Sobol’ indices to be zero because there is no contribution of the single variable and the combination of the variables to the outcome.

Sobol’ indices obtained using Monte Carlo estimation with different sample sizes are shown in Table \ref{Table:Sobol_Analytical_and_Numerical_Sobol_indices}. Latin hypercube sampling scheme was applied when generated the input set for each sample size. One can see that when the sample size is small, in this case 100 and 250 samples, the estimated Sobol’ indices provide wrong index value and wrong order of contribution. For example, the results with 100 samples show that $x_3$ is the most dominant variable while it supposed to have no contribution. While the results using 250 samples show the domination of $S_2$ and $S_{23}$, the index of $S_2$ is shown larger than that of $S_23$ which order is incorrect. From 500 samples onward, the indices show quite consistent results where $S_2$ and $S_{23}$ dominating other variable’s indices and also have the right order of contribution. The index values of $S_2$ and $S_{23}$ are observed to be close enough to the analytical results with small variation. The small discrepancy may come from the nature of the Monte Carlo integration. For the very small indices, it is observed that some values are negative. The negative values may come from the inherent error in Monte Carlo integration as the Sobol’ indices are not supposed to have negative value. However, the absolute values of those negative indices are very small indicating the contribution of the variables are negligible and, thus, can be ignored.

\begin{table}[H]
  \caption{Analytically and numerically obtained Sobol' indices
   \label{Table:Sobol_Analytical_and_Numerical_Sobol_indices}}
  \begin{tabular}{|c|c|c|c|c|c|c|c|c|} \hline
    Sobol' & Analytically & \multicolumn{7}{c|}{Monte Carlo estimated index with sample size of}\\ \cline{3-9}   
    indices & obtained index & $100$ & $250$ & $500$ & $1000$ & $2000$ & $3000$ & $4000$\\ \hline
    $S_{1}$ & $0.0005$ & $-0.0115$ & $0.0742$ & $-0.0266$ & $0.0106$ & $-0.0114$ & $-0.0282$ & $-0.0212$ \\
    $S_{2}$ & $0.4281$ & $0.4556$ & $0.5524$ & $0.4375$ & $0.4432$	& $0.4538$ & $0.4281$ & $0.4383$ \\
    $S_{3}$ & $0.0000$ & $0.6126$ & $0.0547$ & $-0.1432	$ & $0.0418$ & $-0.0176$ & $0.0165$ & $-0.1024$\\
    $S_{12}$ & $0.0007$ & $0.0113$ & $-0.0705$ & $0.0278$ & $-0.0105$ & $0.0131$ & $0.0268$ & $0.0234$ \\
    $S_{13}$ & $0.0000$ & $0.0140$ & $-0.0751$ & $0.0292$ & $-0.0098$ & $0.0107$ & $0.0287$ & $0.0219$\\
    $S_{23}$ & $0.5708$ & $-0.0679$ & $0.3892$ & $0.7045$ & $0.5149$ & $0.5620$ & $0.5568$	& $0.6619$\\
    $S_{123}$ & $0.0000$ & $-0.0140$ & $0.0751$ & $-0.0292$ & $0.0098$ & $-0.0107$ & $-0.0287$ & $-0.0219$\\
    \hline 
  \end{tabular}
\end{table}

\begin{figure}
  \centering
  \subfigure[100 samples]{
    \includegraphics[scale=0.5]{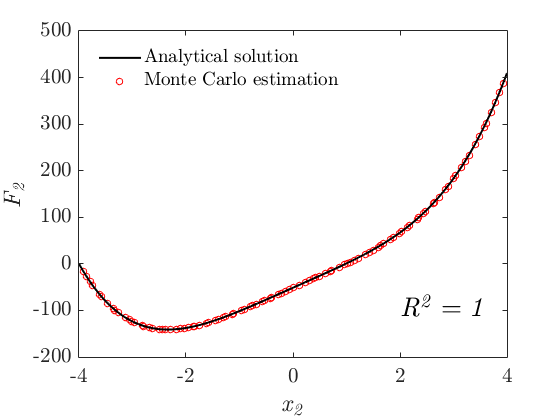}}
  \subfigure[500 samples]{
    \includegraphics[scale=0.5]{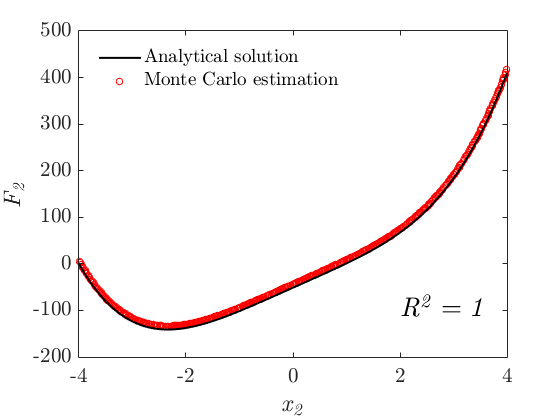}}
  \subfigure[2000 samples]{
    \includegraphics[scale=0.5]{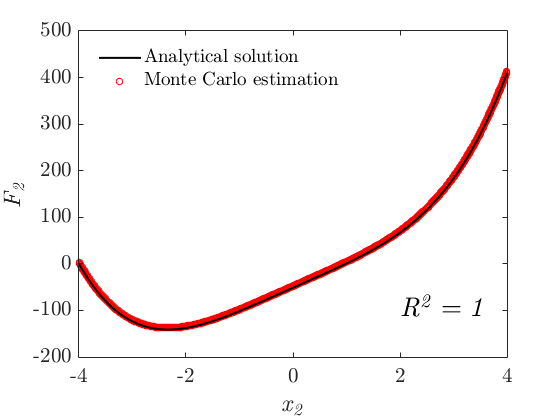}}
  \subfigure[4000 samples]{
    \includegraphics[scale=0.5]{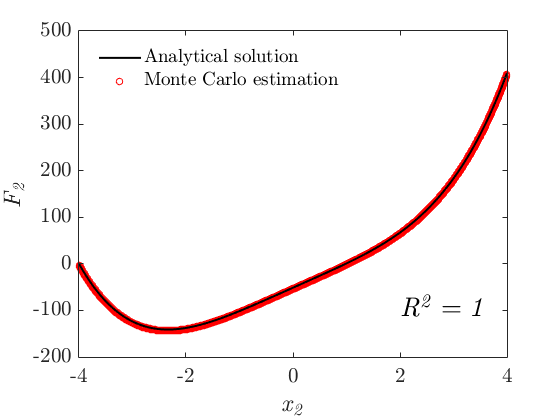}}
  \caption{Comparison of analytical solution
    and Monte Carlo estimation for Sobol'
    function $F_{2}$
[Note: Monte Carlo estimation can estimate Sobol’ functions very close to the analytical solutions ($R^2$=1). However, small samples may lead to erroneous Sobol’ variance calculation]
.\label{Fig1:Sobol_MC_F2}}
\end{figure}

\begin{figure}
  \centering
  \subfigure[100 samples]{
    \label{Fig2:Sobol_MC_F23_a}
    \includegraphics[scale=0.5]{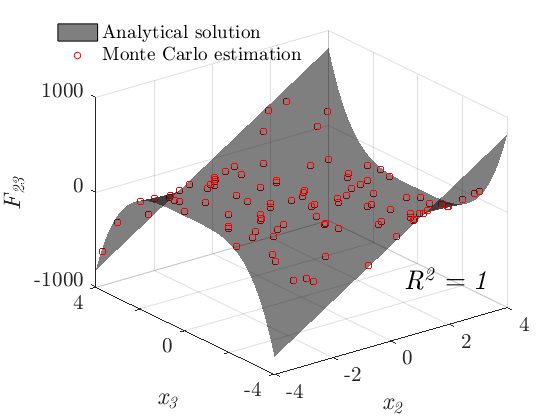}}
  \subfigure[500 samples]{
    \includegraphics[scale=0.5]{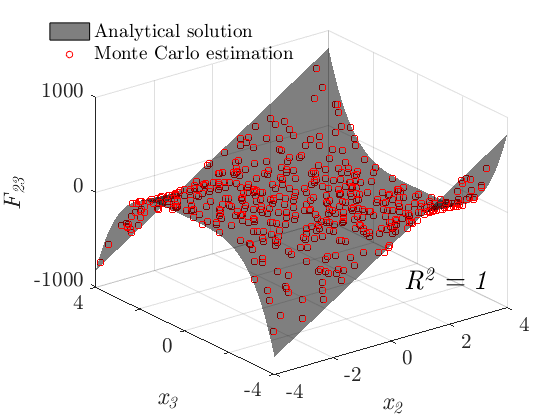}}
  \subfigure[2000 samples]{
    \includegraphics[scale=0.5]{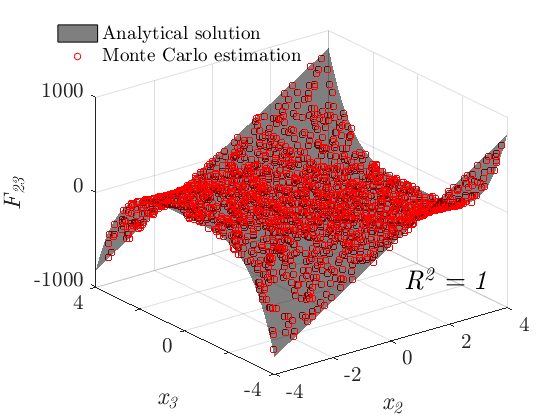}}
  \subfigure[4000 samples]{
    \includegraphics[scale=0.5]{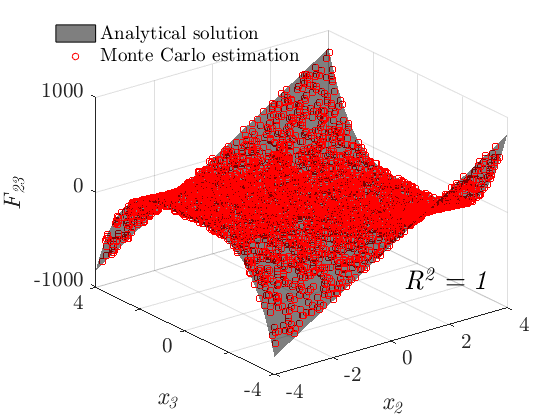}}
  \caption{Comparison of analytical solution
    and Monte Carlo estimation for Sobol'
    function $F_{23}$
[Note: Monte Carlo estimation can estimate Sobol’ functions very close to the analytical solutions ($R^2$=1). However, small samples may lead to erroneous Sobol’ variance calculation]
.\label{Fig2:Sobol_MC_F23}}
\end{figure}

Figures \ref{Fig1:Sobol_MC_F2} and \ref{Fig2:Sobol_MC_F23} show the comparison of estimated Sobol’ functions $F_2$ and $F_{23}$ in comparison to the analytically obtained functions for dominant variables $x_2$ and combination variables $(x_2,x_3)$. It is observed that even with small number of sampling, the estimated function is still very close to the analytical solutions as indicated by the coefficient of determination $R^2$ equals to 1. Although the Sobol’ functions estimation is very accurate, the small number of sample size cannot fully populate input space regions. The incomplete population, such as that with 100 samples shown in Figure \ref{Fig2:Sobol_MC_F23_a}, makes the estimation the variances to be erroneous leading to the wrong Sobol’ indices as shown in Table \ref{Table:Sobol_Analytical_and_Numerical_Sobol_indices}.

After all, once the converged indices is obtained, one can generate a new model function with reduced dimensions of the original mathematical model by selecting only the Sobol’ functions that dominantly contribute to the outcome. Depending on the desired accuracy, one may choose to include as many Sobol’ functions as needed. In this problem, based on the analytically obtained Sobol’ index values, the summation of $S_2$ and $S_{23}$ is 0.999. Therefore, the following new function containing only $x_2$ and combination $(x_2,x_3)$ can have 99.9\% accuracy of the original mathematical model: 

\begin{align}
  \label{Eqn:Sobol_reduced_function}
  F(x_1,x_2,x_3) \approx F(x_2,x_3) = F_0+F_2(x_2)+F_{23}(x_2,x_3)
\end{align}
This advantage will be used to generate new design
equations for 1D phononic crystals presented in the
next sections.

\section{APPLICATION ON 1D PHONONIC CRYSTAL}
\label{Sec:S3_Sobol_on_phononic}
\begin{figure}
  \centering
  \includegraphics[scale=1]{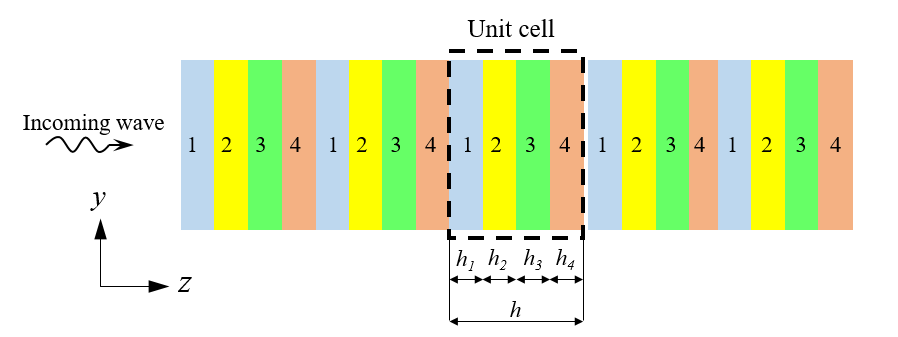}
  \caption{Schematic of 1D phononic crystal.
    \label{Fig3:Sobol_Schematic_1D_phononic}}
\end{figure}

In 1D phononic crystals, the crystal lattice is repeated only in one direction as illustrated in Figure \ref{Fig3:Sobol_Schematic_1D_phononic}. With this arrangement, the 1D phononic crystals are effective in isolating incoming waves with the direction of propagation normal to the crystal lattice. In theory, the phononic crystals consist of infinite number of unit cells. A unit cell defines the smallest group of layers that is periodically repeated in the crystal structures. The unit cell generally consists of multiple layers ranging from a minimum of two layers to any desired number of layers. For the illustration purpose, the unit cell of the 1D phononic crystal depicted in Figure \ref{Fig3:Sobol_Schematic_1D_phononic} is composed of four layers. In the study case presented later, we consider the simplest unit cell that consist of two layers.

This section presents the application of Sobol’ sensitivity analysis on 1D phononic crystals to characterize the influential parameters in obtaining the frequency band gaps. In this study, we focus the attention toward the first frequency band gap subjected to each of S-Wave and P-Wave as the objective functions. Therefore, a theoretical derivation of the dispersion relation in the 1D phononic crystals to obtain the frequency band gaps is first presented.

\subsection{Theory of 1D phononic crystals}
To study the property of phononic crystals, such as frequency band gaps, one can analyze the structures by considering only a single unit cell. This section presents the formulation to obtain the frequency band gaps in the 1D phononic crystals subjected to S-Wave and P-Wave based on transfer matrix method.

\begin{figure}
  \centering
  \includegraphics[scale=1.25]{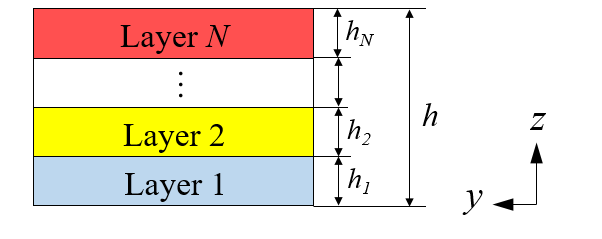}
  \caption{A unit cell with $N$ layers.
    \label{Fig4:Sobol_unit_cell}}
\end{figure}

Suppose the unit cell is composed of $N$ layers stacked in $z$ direction, as shown in Figure \ref{Fig4:Sobol_unit_cell}. The wave oscillation in each layer $n$ can be expressed using the elastic wave equation shown in Eq.~\eqref{Eqn:Sobol_wave_equation}.

\begin{align}
  \label{Eqn:Sobol_wave_equation}
  \dfrac{\partial ^2u_n}{\partial t^2}=C^2_n\dfrac{\partial ^2u_n}{\partial z_n^2}
\end{align}

where $u_n$ is the displacement at layer $n$ and $z_n$ represents the position of interest within the layer with reference to the bottom of each layer $n$. The speed constant $C_n$ for each of S-Wave and P-Wave is shown in Eqs. ~\eqref{Eqn:Sobol_C_SWave} and ~\eqref{Eqn:Sobol_C_PWave}, respectively.

\begin{align}
  \label{Eqn:Sobol_C_SWave}
  C_n=\sqrt{\mu_n/\rho_n}
\end{align}
\begin{align}
  \label{Eqn:Sobol_C_PWave}
  C_n=\sqrt{(\lambda_n+2\mu_n)/\rho_n}
\end{align}
where $\lambda_n$ and $\mu_n$ are, respectively, the first and second Lam{\'e} parameter constants and $\rho_n$ is the material density at layer $n$. 

Since phononic crystals can be applied to vastly different scales, it is more convenient to analyze the property of the unit cell in non-dimensional manner. Therefore, the formulation for dispersion relationship to obtain frequency band gaps is carried out in non-dimensional forms. The reference variables for the characteristic quantities of mass $(M_*)$, length $(L_*)$ and time $(T_*)$ are selected from the power laws combination of total height or thickness of the unit cell $h$ and density $(\rho_1)$ as well as Young’s modulus $(E_1)$ of the first layer, as shown in Eq.~\eqref{Eqn:Sobol_reference_variables}.

\begin{align}
  \label{Eqn:Sobol_reference_variables}
   \begin{split}
   M_*=\rho_1h^3 \\
    L_*=h \\
   T_*=h \sqrt{\rho_1/E_1}
   \end{split}
\end{align}

One can obtain the dimensionless variables by scaling the dimensional quantities with the reference variables. The dimensionless variables used in the derivation of dispersion relationship are summarized in the APPENDIX section. Substitution of dimensionless variables into the original wave equation [Eq.~\eqref{Eqn:Sobol_wave_equation}] gives the dimensionless wave equation [Eq.~\eqref{Eqn:Sobol_wave_equation_nondim}], which retain the characteristic properties of the original wave equation.

\begin{align}
  \label{Eqn:Sobol_wave_equation_nondim}
  \dfrac{\partial ^2\hat{u}_n}{\partial \hat{t}^2}=\hat{C}_n^2\dfrac{\partial ^2\hat{u}_n}{\partial \hat{z}_n^2}
\end{align}

The general solution of Eq.~\eqref{Eqn:Sobol_wave_equation_nondim} is shown in Eq~\eqref{Eqn:Sobol_wave_equation_general_solution}. Substitution of Eq.~\eqref{Eqn:Sobol_wave_equation_general_solution} to Eq.\eqref{Eqn:Sobol_wave_equation_nondim} yields Eq.~\eqref{Eqn:Sobol_spatial_wave_equation}. The steady-state displacement that satisfies Eq.~\eqref{Eqn:Sobol_spatial_wave_equation} is shown in Eq.~\eqref{Eqn:Sobol_steady_state_displacement}.

\begin{align}
  \label{Eqn:Sobol_wave_equation_general_solution}
  \hat{u}_n(\hat{z}_n,\hat{t})=e^{i \hat{\omega t}}\hat{u}_n(\hat{z}_n)
\end{align}
%
\begin{align}
  \label{Eqn:Sobol_spatial_wave_equation}
  \hat{C}_n^2\dfrac{\partial^2\hat{u}_n(\hat{z}_n)}{\partial \hat{z}_n^2}+\hat{\omega}^2\hat{u}_n(\hat{z}_n)=0
\end{align}
%
\begin{align}
  \label{Eqn:Sobol_steady_state_displacement}
 \hat{u}_n(\hat{z}_n)=A_n \text{sin}(\hat{\omega}\hat{z}_n/\hat{C}_n)+B_n \text{cos}(\hat{\omega}\hat{z}_n/\hat{C}_n)
\end{align}

The terms $A_n$ and $B_n$ in Eq.~\eqref{Eqn:Sobol_steady_state_displacement} are the amplitudes of the steady state displacement solution on layer $n$. In the elastic body of each layer $n$, the constitutive equations for normal and shear stresses are shown in Eqs.~\eqref{Eqn:Sobol_normal_stress} and ~\eqref{Eqn:Sobol_shear_stress}, respectively. To obtain the dispersion curve of the S-Wave, Eqs.~\eqref{Eqn:Sobol_steady_state_displacement} and ~\eqref{Eqn:Sobol_shear_stress} are arranged into matrix form, as shown in Eq.~\eqref{Eqn:Sobol_disp_stress_matrix}, with the $\hat{C_n}$ constant for the S-Wave propagation.

\begin{align}
   \begin{split}
  \label{Eqn:Sobol_normal_stress}
  \hat{\sigma}_n(\hat{z_n})=(\hat{\lambda}_n+2\hat{\mu}_n)\partial \hat{u}_n/\partial \hat{z}_n\\
  =(\hat{\lambda}_n+2\hat{\mu}_n)\hat{\omega}[A_n \text{cos}(\hat{\omega}\hat{z}_n/\hat{C}_n)-B_n \text{sin}(\hat{\omega}\hat{z}_n/\hat{C}_n)]/\hat{C}_n
   \end{split}
\end{align}

\begin{align}
   \begin{split}
  \label{Eqn:Sobol_shear_stress}
  \hat{\tau}_n(\hat{z_n})=\hat{\mu}_n \partial \hat{u}_n/\partial \hat{z}_n\\
  =\hat{\mu}_n \hat{\omega}[A_n \text{cos}(\hat{\omega}\hat{z}_n/\hat{C}_n)-B_n \text{sin}(\hat{\omega}\hat{z}_n/\hat{C}_n)]/\hat{C}_n
   \end{split}
\end{align}

\begin{align}
   \begin{split}
  \label{Eqn:Sobol_disp_stress_matrix}
  \begin{Bmatrix}  \hat{u}_n(\hat{z}_n) \\ \hat{\tau}_n(\hat{z}_n)  \end{Bmatrix}=
  \begin{bmatrix}  sin(\hat{\omega}\hat{z}_n/\hat{C}_n) & cos(\hat{\omega}\hat{z}_n/\hat{C}_n)\\ 
                               \dfrac{\hat{\mu}_n\hat{\omega}}{\hat{C}_n}cos(\hat{\omega}\hat{z}_n/\hat{C}_n)  & -\dfrac{\hat{\mu}_n\hat{\omega}}{\hat{C}_n}sin(\hat{\omega}\hat{z}_n/\hat{C}_n)\end{bmatrix}
  \begin{Bmatrix}  A_n \\ B_n  \end{Bmatrix}\\
  \text{or}  \hspace{1cm} \hat{\textbf{w}}_n(\hat{z}_n)=\hat{\textbf{H}}_n(\hat{z}_n)\mathbf{\Psi}_n
  \end{split}
\end{align}

The left-hand side vector of Eq.~\eqref{Eqn:Sobol_disp_stress_matrix} at the bottom of layer $n$ is defined as $\hat{\textbf{w}}_n^b$  [Eq.~\eqref{Eqn:Sobol_disp_stress_bottom_n}], which gives information regarding the displacement and the stress at the bottom of layer $n$. As for the top of layer $n$, the left-hand side vector is defined as $\hat{\textbf{w}}_n^t$  [Eq.~\eqref{Eqn:Sobol_disp_stress_top_n}], which gives the information of displacement and the stress at the top of layer $n$. Eq.~\eqref{Eqn:Sobol_disp_stress_bottom_n} can be related to Eq.~\eqref{Eqn:Sobol_disp_stress_top_n} through a transfer matrix $\hat{\textbf{T}}_n$, as shown in Eq.~\eqref{Eqn:Sobol_disp_stress_top_bottom_n}. Hence, the transfer matrix $\hat{\textbf{T}}_n$ for a single layer $n$ is denoted in Eq.~\eqref{Eqn:Sobol_T_matrix_n}

\begin{align}
  \label{Eqn:Sobol_disp_stress_bottom_n}
  \hat{\mathbf{w}}_n^b \equiv \hat{\mathbf{w}}_n(0)=\hat{\textbf{H}}_n(0)\mathbf{\Psi}_n
\end{align}
%
\begin{align}
  \label{Eqn:Sobol_disp_stress_top_n}
  \hat{\textbf{w}}_n^t \equiv \hat{\textbf{w}}_n(\hat{h}_n)=\hat{\textbf{H}}_n(\hat{h}_n)\mathbf{\Psi}_n
\end{align}
%
\begin{align}
  \label{Eqn:Sobol_disp_stress_top_bottom_n}
  \hat{\textbf{w}}_n^t=\hat{\textbf{T}}_n \hat{\textbf{w}}_n^b
\end{align}
%
\begin{align}
  \label{Eqn:Sobol_T_matrix_n}
  \hat{\textbf{T}}_n=\hat{\textbf{H}}_n(\hat{h}_n)[\hat{\textbf{H}}_n(0)]^{-1}
\end{align}

Each layer’s interface of the unit cell is assumed to be perfectly bonded; hence the displacement and shear stress need to satisfy continuity. Therefore, the displacement and shear stress of the top of layer $n$ are equal to that of the bottom of layer $n+1$, as indicated in Eq.~\eqref{Eqn:Sobol_continuity}. Subsequently, the relationship of displacement and shear stress of the bottom and top surfaces of the unit cell containing the $N$ layers can be expressed as in Eq.~\eqref{Eqn:Sobol_disp_stress_top_bottom_unit_cell}. By changing the displacement and shear stress vector of the top and bottom surface of the unit cell as $\hat{\textbf{w}}^t=\hat{\textbf{w}}^t_N$ and $\hat{\textbf{w}}^b=\hat{\textbf{w}}^b_1$, respectively, Eq.~\eqref{Eqn:Sobol_disp_stress_top_bottom_unit_cell} can be shortened into Eq.~\eqref{Eqn:Sobol_T_matrix_unit_cell}.

\begin{align}
  \label{Eqn:Sobol_continuity}
  \hat{\textbf{w}}_{n+1}^b=\hat{\textbf{w}}_{n}^t
\end{align}
\begin{align}
  \label{Eqn:Sobol_disp_stress_top_bottom_unit_cell}
  \hat{\textbf{w}}_N^t=\hat{\textbf{T}}_N \hat{\textbf{w}}_N^b
   =\hat{\textbf{T}}_N \hat{\textbf{w}}_{N-1}^t
   =\hat{\textbf{T}}_N \hat{\textbf{T}}_{N-1} \hat{\textbf{w}}_{N-1}^b
   =\cdots
   =(\hat{\textbf{T}}_N \hat{\textbf{T}}_{N-1} \cdots \hat{\textbf{T}}_1)\hat{\textbf{w}}_{1}^b
\end{align}
\begin{align}
  \label{Eqn:Sobol_T_matrix_unit_cell}
  \hat{\textbf{w}}^t=\hat{\textbf{T}}(\hat{\omega}) \hat{\textbf{w}}^b
\end{align}

The transfer matrix for a unit cell of the 1D phononic crystals is $\hat{\textbf{T}}(\hat{\omega})=\hat{\textbf{T}}_N \hat{\textbf{T}}_{N-1} \cdots \hat{\textbf{T}}_1$. Based on the Bloch-Floquent theorem, the periodic boundary conditions can be expressed as in Eq.~\eqref{Eqn:Sobol_periodic_BC}, in which $\hat{h}$ is the dimensionless thickness of unit cell, $\hat{k}$ is the dimensionless wavenumber in reciprocal lattice space and $i$ is the imaginary number. Subtraction of Eq.~\eqref{Eqn:Sobol_periodic_BC} from Eq.~\eqref{Eqn:Sobol_T_matrix_unit_cell} yields Eq.~\eqref{Eqn:Sobol_T_matrix_and_periodic_BC} and the nontrivial solution can be achieved when the determinant is equal to zero, as shown in Eq.~\eqref{Eqn:Sobol_eigen}.

\begin{align}
  \label{Eqn:Sobol_periodic_BC}
  \hat{\textbf{w}}^t=e^{i\hat{k}\hat{h}} \hat{\textbf{w}}^b
\end{align}
%
\begin{align}
  \label{Eqn:Sobol_T_matrix_and_periodic_BC}
  [\hat{\textbf{T}}(\hat{\omega})-e^{i\hat{k}\hat{h}}\textbf{I}]\hat{\textbf{w}}^b=\mathbf{0}
\end{align}
%
\begin{align}
  \label{Eqn:Sobol_eigen}
  |\hat{\textbf{T}}(\hat{\omega})-e^{i\hat{k}\hat{h}}\textbf{I}|=\mathbf{0}
\end{align}

Eq.~\eqref{Eqn:Sobol_eigen} is the so-called Eigenvalue problem, with $e^{i\hat{k}\hat{h}}$ equal to the Eigenvalue of the transfer matrix $\hat{\textbf{T}}(\hat{\omega})$. Thus the relationship between wavenumber $\hat{k}$ and frequency $\hat{\omega}$  can be obtained by solving the corresponding Eigenvalue problem. The relationship between the wavenumber and frequency forms the S-Wave dispersion curve. The curves are related to real wave-numbers and the frequency band gaps are related to complex wave-numbers. Although the wavenumber $\hat{k}$ is unrestricted, it is only necessary to consider $\hat{k}$ limited to the first Brillouin zone \citep{kittel1996introduction}, i.e. $\hat{k}\in[-\pi/\hat{h},\pi/\hat{h}]$, to obtain the frequency band gaps. Likewise, the P-Wave dispersion curve can be obtained through a similar approach by arranging Eqs.~\eqref{Eqn:Sobol_steady_state_displacement} and ~\eqref{Eqn:Sobol_normal_stress} into the matrix form of Eq.~\eqref{Eqn:Sobol_disp_stress_matrix} and using the $\hat{C}_n$ constant for the P-Wave.

Although Eq.~\eqref{Eqn:Sobol_eigen} can be solved directly, one can further simplify the form of the equation. It is known that the transfer matrix $\hat{\textbf{T}}(\hat{\omega})$ is a two by two matrix. Hence, the Eigenvalues $L$ of the $\hat{\textbf{T}}(\hat{\omega})$ matrix can be obtained using Cayley-Hamilton theorem, as shown in Eq.~\eqref{Eqn:Sobol_Cayley_Hamilton}
%
\begin{align}
  \label{Eqn:Sobol_Cayley_Hamilton}
  L^2-I_1(\hat{\omega})L+I_2(\hat{\omega})=0
\end{align}
where $I_1(\hat{\omega})=\textbf{tr}(\hat{\textbf{T}}(\hat{\omega}))=\hat{\text{T}}_{11}+\hat{\text{T}}_{22}$ and $I_2(\hat{\omega})=|\hat{\textbf{T}}(\hat{\omega})|$ are the first and second variants of the transformation matrix, respectively. From Eq.~\eqref{Eqn:Sobol_T_matrix_n}, the determinant of transfer matrix at each layer $n$ is
%
\begin{align}
  \label{Eqn:Sobol_determinant_T_matrix_n}
  |\hat{\textbf{T}}_n|=|\hat{\textbf{H}}_n(\hat{h}_n)||\hat{\textbf{H}}_n(0)|^{-1}
                                    =\left(\dfrac{-\hat{\mu}\hat{\omega}}{\hat{C}_n}\right)\left(\dfrac{\hat{C}_n}{-\hat{\mu}\hat{\omega}}\right)=1
\end{align}
Hence the second invariant becomes $I_2(\hat{\omega})=|\hat{\textbf{T}}(\hat{\omega})|=|\hat{\textbf{T}}_N||\hat{\textbf{T}}_{N-1}|\cdots|\hat{\textbf{T}}_1|=1$. The polynomial equation shown in Eq.~\eqref{Eqn:Sobol_Cayley_Hamilton} can be written as
%
\begin{align}
  \label{Eqn:Sobol_Cayley_Hamilton_1st_invariant}
   L+L^{-1}=I_1(\hat{\omega})
\end{align}

Note that $L+L^{-1}=e^{i\hat{k}\hat{h}}+e^{-i\hat{k}\hat{h}}=2\text{cos}(\hat{k}\hat{h})$. Therefore, a simpler form of dispersion relationship is obtained.

\begin{align}
  \label{Eqn:Sobol_simpler_1st_invariant}
   \text{cos}(\hat{k}\hat{h})=\dfrac{1}{2}I_1(\hat{\omega})
\end{align}

In this paper, we consider the simplest form of 1D phononic crystals, which unit cells consist of two layers. Thus, Eq.~\eqref{Eqn:Sobol_simpler_1st_invariant} can be expanded into Eq.~\eqref{Eqn:Sobol_2layer_dispersion} that works for both S-Wave and P-Wave.

\begin{align}
  \label{Eqn:Sobol_2layer_dispersion}
   \text{cos}(\hat{k}\hat{h})=\text{cos}\left(\dfrac{\hat{\omega}\hat{h}_1}{\hat{C}_1}\right)
                                                 \text{cos}\left(\dfrac{\hat{\omega}\hat{h}_2}{\hat{C}_2}\right)
 				      -\dfrac{1}{2}\left(\dfrac{\hat{\rho}_1\hat{C}_1}{\hat{\rho}_2\hat{C}_2}+\dfrac{\hat{\rho}_2\hat{C}_2}{\hat{\rho}_1\hat{C}_1}\right)
				      \text{sin}\left(\dfrac{\hat{\omega}\hat{h}_1}{\hat{C}_1}\right)
                                                 \text{sin}\left(\dfrac{\hat{\omega}\hat{h}_2}{\hat{C}_2}\right)
\end{align}

\begin{figure}
  \centering
  \subfigure[S-Wave]{
    \includegraphics[scale=0.5]{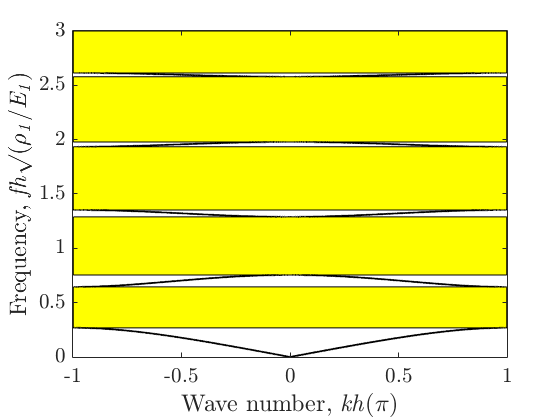}}
  \subfigure[P-Wave]{
    \includegraphics[scale=0.5]{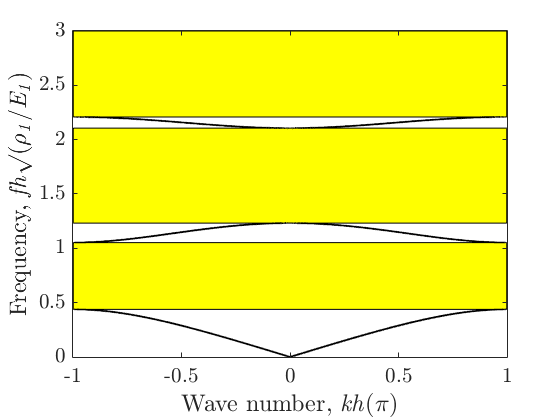}}
  \caption{Typical dispersion curves.
    \label{Fig5:Sobol_dispersion}}
\end{figure}

Figure \ref{Fig5:Sobol_dispersion} shows the typical dispersion curves of 1D phononic crystals subjected to each of S-Wave and P-Wave. The dispersion curves were constructed from a unit cell consisting of two different materials with the following properties, $E_2/E_1=1000$, $\rho_2/\rho_1=2$, $h_2/h_1=2$ and $\upsilon_2=\upsilon_1=0.2$. The yellow hatched areas indicate the frequency band gaps, in which the wave propagation is forbidden. In general, the frequency band gaps of S-Wave is located at lower frequency because it has a lower wave speed than P-Wave. 

\subsection{Sensitivity analysis on 1D phononic crystals}
Sobol' sensitivity analysis is strongly correlated to the input space. Since we performed our study in non-dimensional form, the input parameters are selected as the ratio of the material and geometric properties between the two layers in the unit cell with the softer and lighter layer set as the reference. For this study, the input parameters were varied as shown in Table \ref{Table:Sobol_sensitivity_parameters}. The parameters were generated using Latin Hypercube sampling scheme. The first three parameters $(E_2/E_1,\rho_2/\rho_1,h_2/h_1)$ are uniformly distributed in the logarithmic scale while the Poisson’s ratios are uniformly distributed in linear scale. The highest value of Poisson’s ratio to be used in this study was selected as 0.463. The reason of such selection is because as the Poisson’s ratio gets closer to 0.5, the first Lam\'{e} parameter goes to infinity. This hypothetical condition is not applicable to any actual material \citep{mott2008bulk} and therefore, is omitted in this study.

Two objective functions are utilized to represent the first frequency band gap, i.e. the starting frequency and the width. These two objective functions are employed to represent the first frequency band gap of each of S-Wave and P-Wave, resulting in four objective functions in total. 

\begin{table}
  \caption{Parameters used in sensitivity analysis
    \label{Table:Sobol_sensitivity_parameters}}
  \begin{tabular}{|c|c|} \hline
    Parameters & Value range \\ \hline
    Young’s modulus ratio $(E_2/E_1)$ & $10-10000$ \\
    Density ratio $(\rho_2/\rho_1)$ & $1-1000$ \\
    Thickness ratio $(h_2/h_1)$ & $0.11-9$ \\
    Poisson’s ratio of material in the first layer $(\upsilon_1)$ & $0-0.463$ \\
    Poisson’s ratio of material in the second layer $(\upsilon_2)$ & $0-0.463$ \\
    \hline 
  \end{tabular}
\end{table}

\begin{figure}
  \centering
  \includegraphics[scale=0.6]{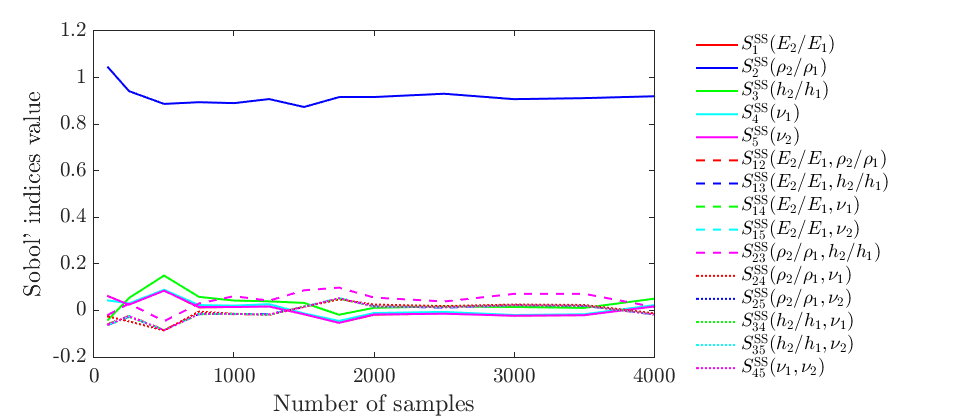}
  \caption{First and second order indices
    for the starting of the first frequency
    band gap of S-Wave.\label{Fig6:Sobol_indices_S-Wave_start}}
\end{figure}

\begin{figure}
  \centering
  \subfigure{
    \includegraphics[scale=0.5]{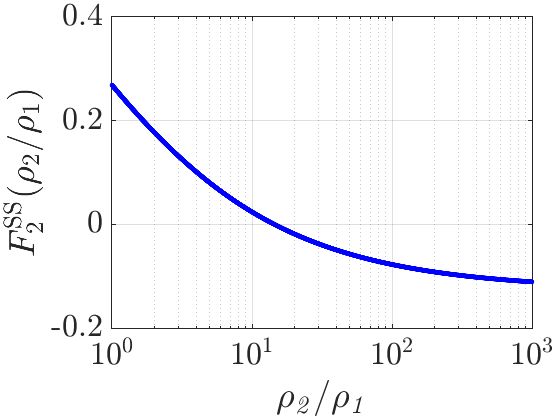}}
  \subfigure{
    \includegraphics[scale=0.5]{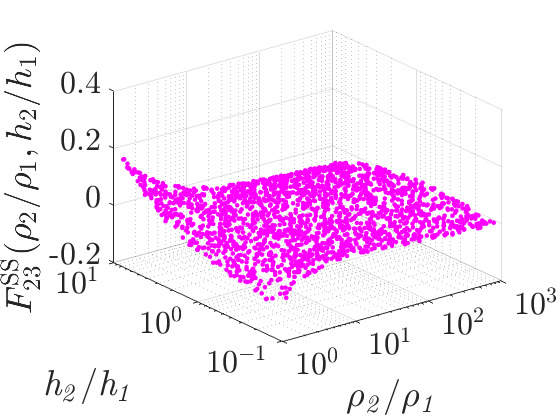}}
  \subfigure{
    \includegraphics[scale=0.5]{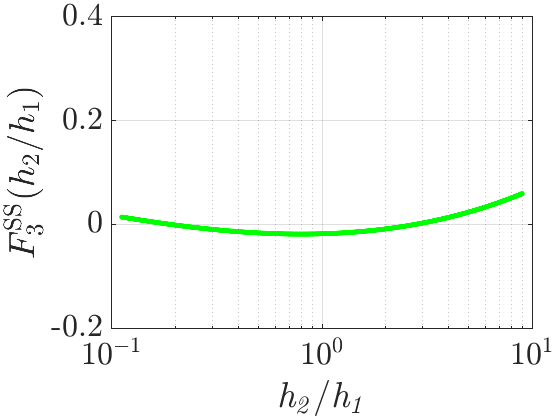}}
  \caption{Influential Sobol’ functions for the starting
    of the first frequency band gap of S-Wave.
    \label{Fig7:Sobol_function_S-Wave_start}}
\end{figure}

Figure \ref{Fig6:Sobol_indices_S-Wave_start} shows the first and second order Sobol’ indices with the starting of the first frequency band gap of S-Wave (denoted by superscript "SS") as the objective function. One can see that the sole dominant parameter that affect the starting of the frequency band gap is the density ratio $(\rho_2/\rho_1)$ of the two layers in the unit cell with Sobol’ index value around 0.9. The remaining parameters and the interaction between the parameters are shown to contribute very little to this objective function. Although not as significant as the density ratio, the second and third highest indices are shown by the combination of parameters $(\rho_2/\rho_1,h_2/h_1)$ and the thickness ratio $(h_2/h_1)$ parameter. The Sobol’ indices value are reflected in their variation of Sobol’ functions (derived from 2000 samples), as shown in Figure \ref{Fig7:Sobol_function_S-Wave_start}. For example, variation of $F_2^{\text{SS}}(\rho_2/\rho_1)$ surrounding its mean value is much larger than $F_{23}^{\text{SS}}(\rho_2/\rho_1,h_2/h_1)$ and $F_3^{\text{SS}}(h_2/h_1)$. In addition, it is observed that larger variation of $F_{23}^{\text{SS}}(\rho_2/\rho_1,h_2/h_1)$ only occurs at the corner where density ratio is small and thickness ratio is high.

\begin{figure}
  \centering
  \includegraphics[scale=0.6]{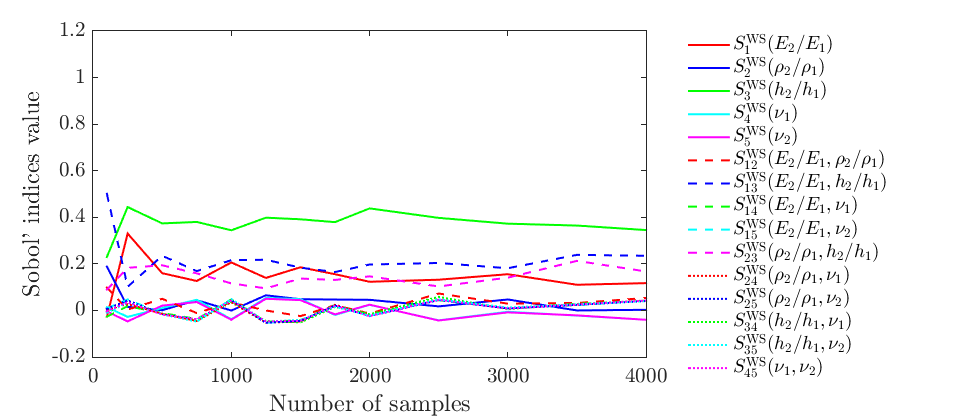}
  \caption{First and second order indices 
    for the width of the first frequency 
    band gap of S-Wave.\label{Fig8:Sobol_indices_S-Wave_width}}
\end{figure}

\begin{figure}
  \centering
  \subfigure{
    \includegraphics[scale=0.5]{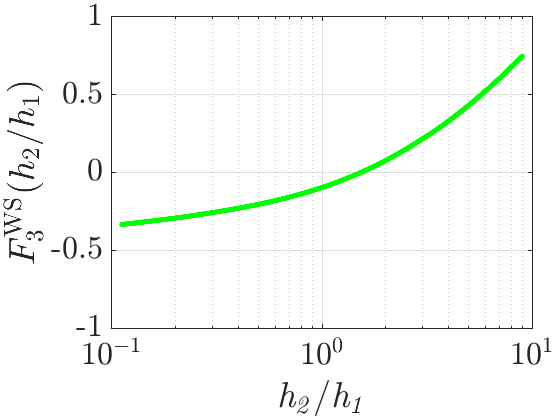}}
  \subfigure{
    \includegraphics[scale=0.5]{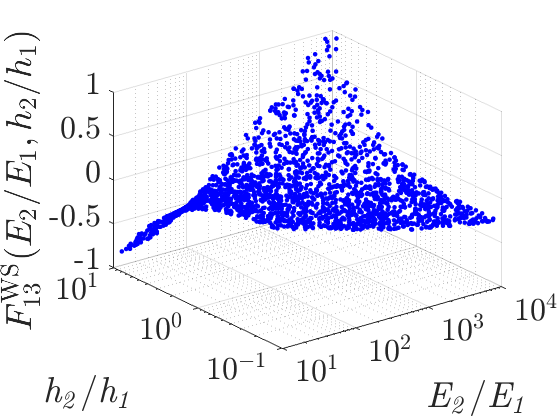}}
  \subfigure{
    \includegraphics[scale=0.5]{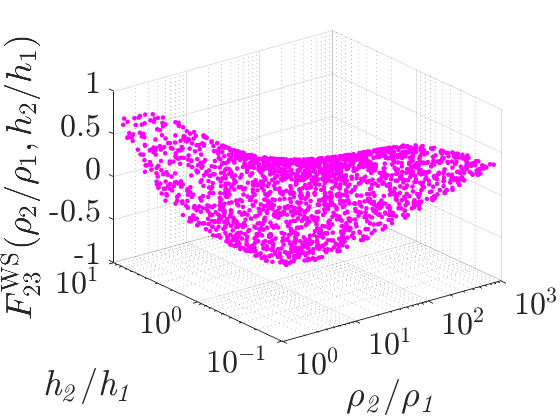}}
  \subfigure{
    \includegraphics[scale=0.5]{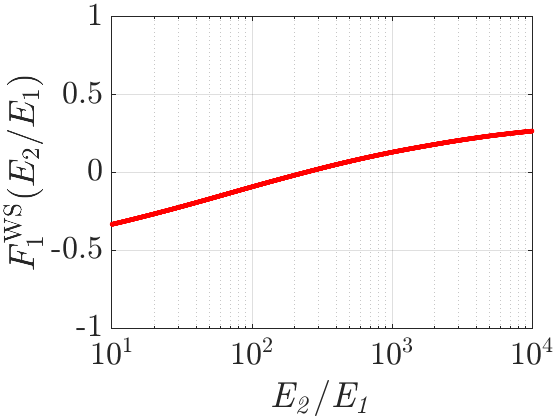}}
  \subfigure{
    \includegraphics[scale=0.5]{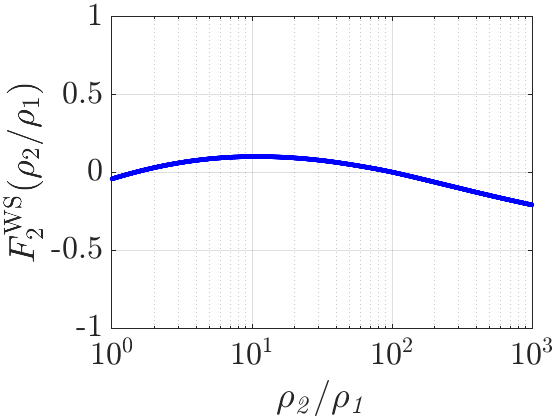}}
  \caption{Influential Sobol’ functions for the width of the first frequency band gap of S-Wave.
    \label{Fig9:Sobol_function_S-Wave_width}}
\end{figure}

On the other hand, the width of the first frequency band gap of S-Wave (denoted by superscript "WS") is affected by more parameters and their interaction. Figure \ref{Fig8:Sobol_indices_S-Wave_width} shows that the highest Sobol’ index is only around 0.4 which is contributed by the thickness ratio $(h_2/h_1)$ parameter. Closely below that, the next three indices that are located between 0.1 to 0.2 index value are contributed by combination of parameters $(E_2/E_1,h_2/h_1)$, $(\rho_2/\rho_1,h_2/h_1)$ and parameter $(E_2/E_1)$. Therefore, the Sobol’ functions associated with the respected parameters and the combination of parameters show a large variation as depicted in Figure \ref{Fig9:Sobol_function_S-Wave_width}. In fact, the variation in Sobol’ functions shown in Figure \ref{Fig9:Sobol_function_S-Wave_width} emphasize the importance to investigate the interaction between parameters involved. For example, one may think that as the thickness ratio gets larger, the width of the first frequency band gap gets wider. The premise is only true when the Young’s modulus ratio is high. As the Young’s modulus ratio decrease, the opposite result is observed. For the parameters with low Sobol’ index value, such as that contributed by $(\rho_2/\rho_1)$, the Sobol’ function varies very little in comparison to other functions.

\begin{figure}
  \centering
  \includegraphics[scale=0.6]{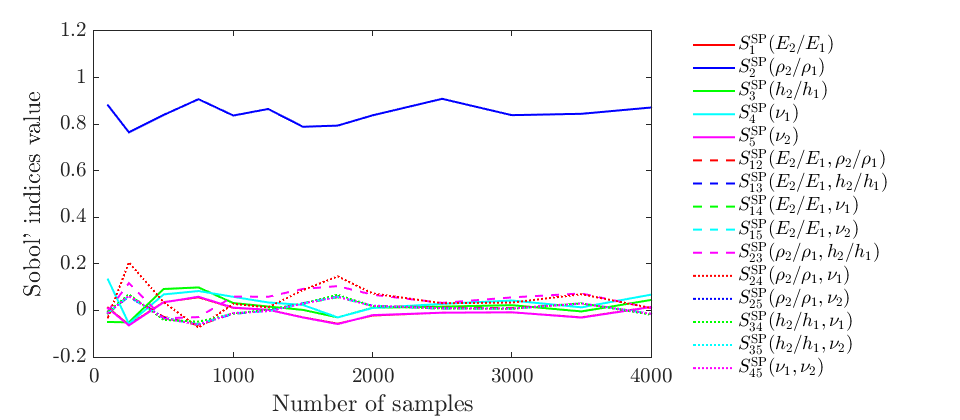}
  \caption{First and second order indices for the
    starting of the first frequency band gap of
    P-Wave.\label{Fig10:Sobol_indices_P-Wave_start}}
\end{figure}

\begin{figure}
  \centering
  \subfigure{
    \includegraphics[scale=0.5]{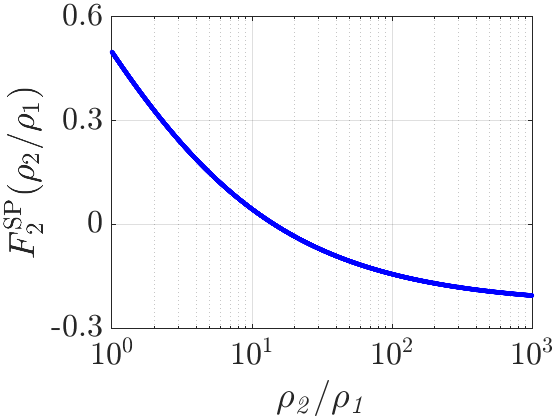}}
  \subfigure{
    \includegraphics[scale=0.5]{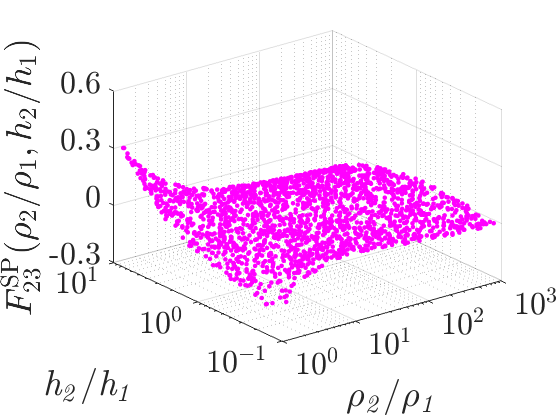}}
  \subfigure{
    \includegraphics[scale=0.5]{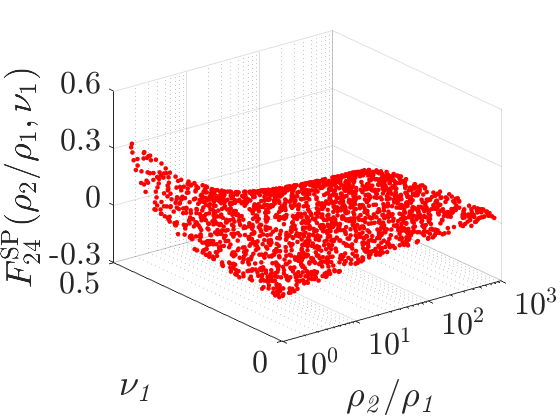}}
  \subfigure{
    \includegraphics[scale=0.5]{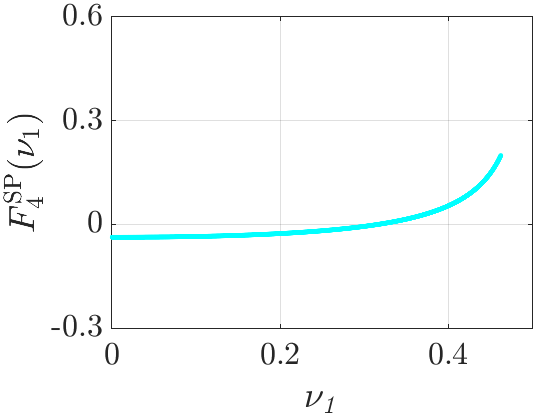}}
  \subfigure{
    \includegraphics[scale=0.5]{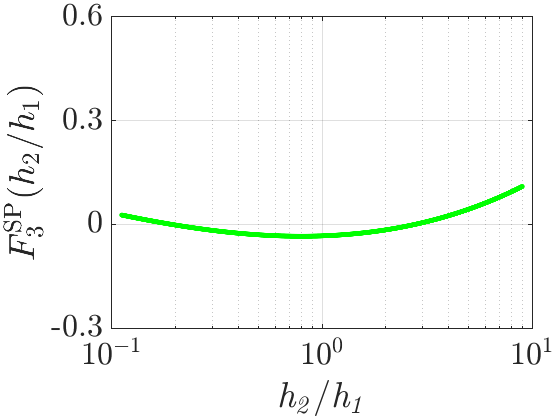}}
  \caption{Influential Sobol’ functions for the starting
    of the first frequency band gap of P-Wave.
    \label{Fig11:Sobol_function_P-Wave_start}}
\end{figure}

For the objective function of the starting of the first frequency band gap of P-Wave (denoted by superscript "SP"), similar results are observed as those of S-Wave. The sole dominant parameter is contributed by the density ratio $(\rho_2/\rho_1)$ parameter with the Sobol’ index showing around 0.8 in Figure \ref{Fig10:Sobol_indices_P-Wave_start}. However, Sobol’ indices only represent the variation of Sobol’ function as a whole. A parameter or a correlation of parameters with small Sobol’ index value may have a Sobol’ function with large variation at very minor portion of the input space. For example, the Poisson’s ratio parameter only starting to affects the frequency band gaps of P-Wave after 0.35. Five most important Sobol’ functions that affects the starting of the first frequency band gaps of P-Wave are shown in Figure \ref{Fig11:Sobol_function_P-Wave_start}.

\begin{figure}
  \centering
  \includegraphics[scale=0.6]{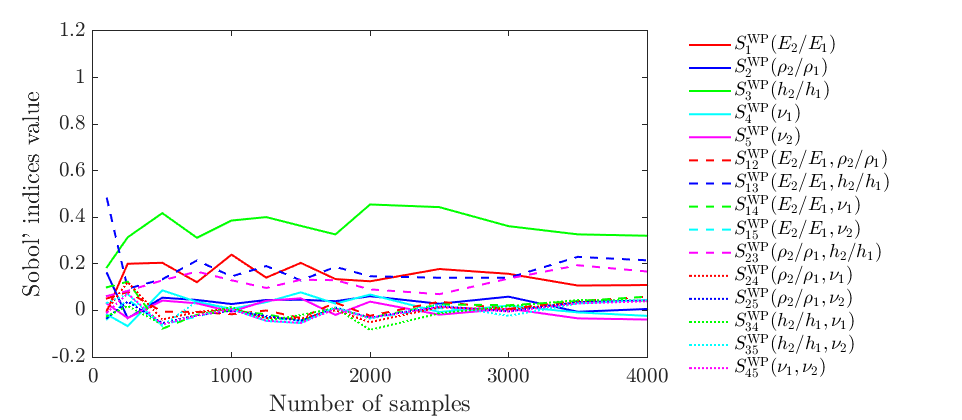}
  \caption{First and second order indices 
    for the width of the first frequency 
    band gap of P-Wave.\label{Fig12:Sobol_indices_P-Wave_width}}
\end{figure}

\begin{figure}
  \centering
  \subfigure{
    \includegraphics[scale=0.5]{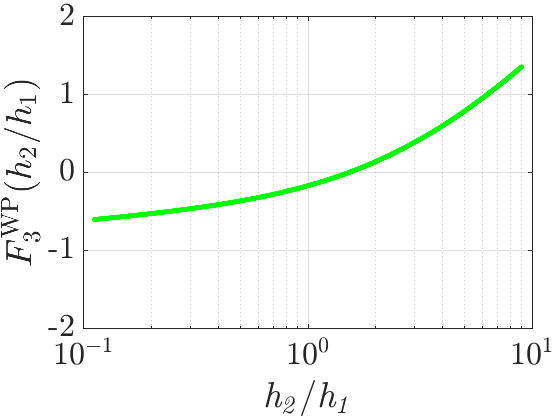}}
  \subfigure{
    \includegraphics[scale=0.5]{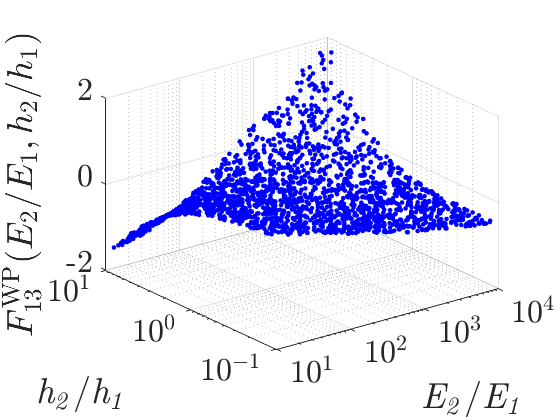}}
  \subfigure{
    \includegraphics[scale=0.5]{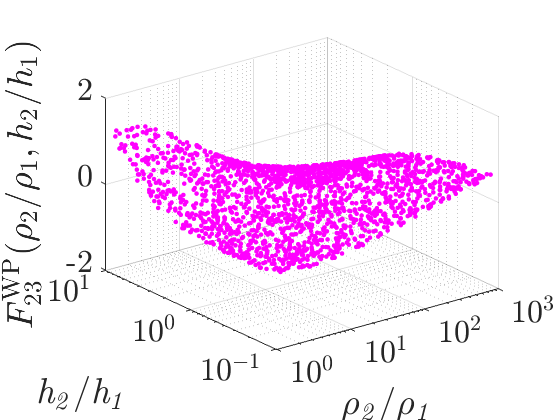}}
  \subfigure{
    \includegraphics[scale=0.5]{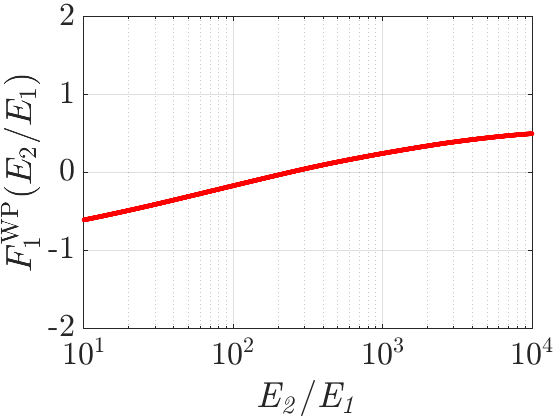}}
  \subfigure{
    \includegraphics[scale=0.5]{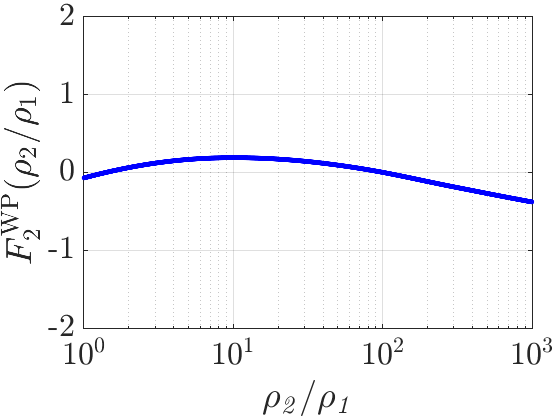}}
  \subfigure{
    \includegraphics[scale=0.5]{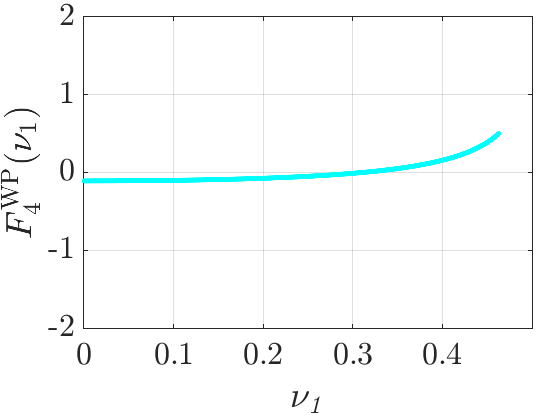}}
  \caption{Influential Sobol’ functions for the width
    of the first frequency band gap of P-Wave.
    \label{Fig13:Sobol_function_P-Wave_width}}
\end{figure}

Similar to that of S-Wave, the objective function for the width of the first frequency band gap of P-Wave  (denoted by superscript "WP") is also affected by more parameters and correlation of parameters. The Sobol’ indices indicate that the thickness ratio $(h_2/h_1)$ is the most influential parameter followed by the correlation of parameters $(E_2/E_1,h_2/h_1)$, $(\rho_2/\rho_1,h_2/h_1)$, individual parameters $(E_2/E_1)$, $(\rho_2/\rho_1)$ and $(\upsilon_1)$ in sequence as depicted  in Figure \ref{Fig12:Sobol_indices_P-Wave_width}. The Sobol’ functions showed in Figure \ref{Fig13:Sobol_function_P-Wave_width} describe how the parameters and the correlation of parameters affect the objective function. In order to get a wide frequency band gap, it is necessary to have a higher thickness ratio, a higher Young’s modulus ratio with a lower density ratio.

\section{DESIGN EQUATIONS BASED ON REDUCED SOBOL' FUNCTION}
\label{Sec:S4_Sobol_phononic_design}

In this study, we exercise the advantage of Sobol’ decomposition that allows the reduction of the dimension of the objective function. As previously mentioned in Section \ref{Sec:S3_Sobol_on_simple_model}, the reduced objective function can be obtained by truncating the series of Sobol’ functions up to the desired level of accuracy according to the Sobol’ indices.

\begin{figure}
  \centering
  \subfigure{
    \includegraphics[scale=0.5]{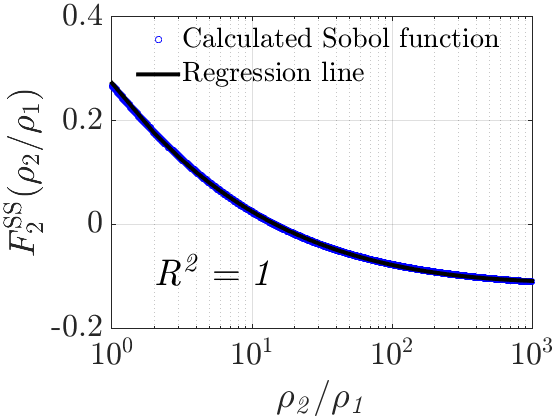}}
  \subfigure{
    \includegraphics[scale=0.5]{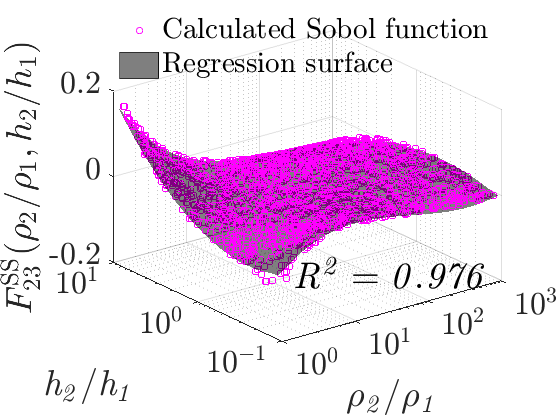}}
  \subfigure{
    \includegraphics[scale=0.5]{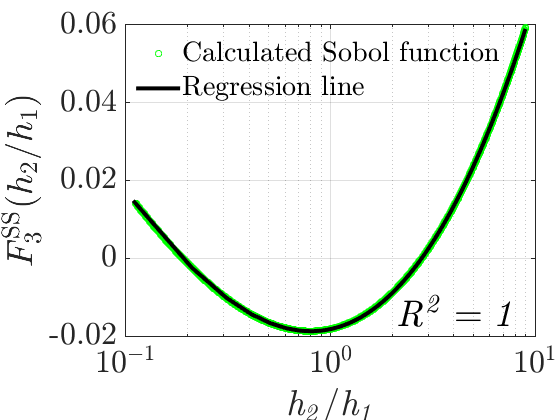}}
  \caption{Curve fitting on Sobol’ functions for
    the starting of the first frequency band gap
    of S-Wave.
    \label{Fig14:Fit_Sobol_function_S-Wave_start}}
\end{figure}

For the objective function of the starting of the first frequency band gap of S-Wave, the summation of the three most influential Sobol’ indices ($S_2^{\text{SS}}[\rho_2/\rho_1]$, $S_{23}^{\text{SS}}[\rho_2/\rho_1,h_2/h_1]$ and $S_3^{\text{SS}}[h_2/h_1]$) is roughly 0.98. Hence, using the summation of the Sobol’ functions associated with those three indices, one can predict the objective function with 98\% confidence. Figure \ref{Fig14:Fit_Sobol_function_S-Wave_start} shows the regression curves and surface for those three Sobol’ functions. An approximate objective function for the starting of the first frequency band gap of S-Wave is shown in Eq.~\eqref{Eqn:Sobol_SWave_start_equation} with the fitted Sobol’ functions tabulated in Table \ref{Table:Sobol_SWave_start_approx_functions}.

\begin{figure}
  \centering
  \subfigure{
    \includegraphics[scale=0.5]{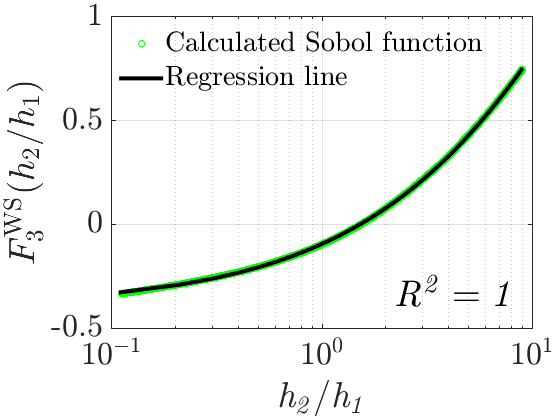}}
  \subfigure{
    \includegraphics[scale=0.5]{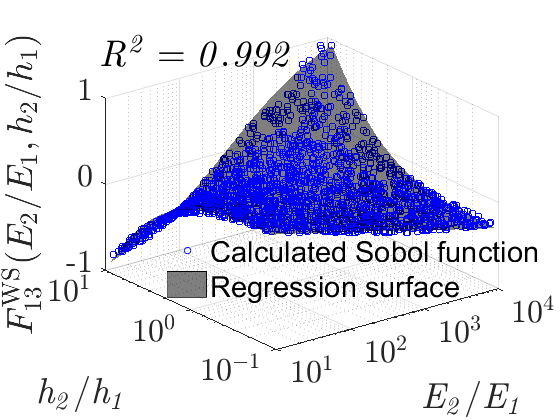}}
  \subfigure{
    \includegraphics[scale=0.5]{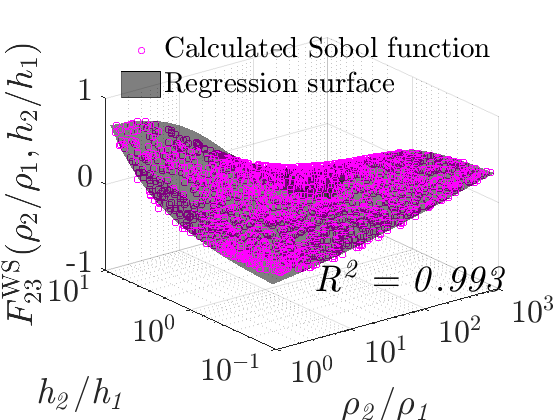}}
  \subfigure{
    \includegraphics[scale=0.5]{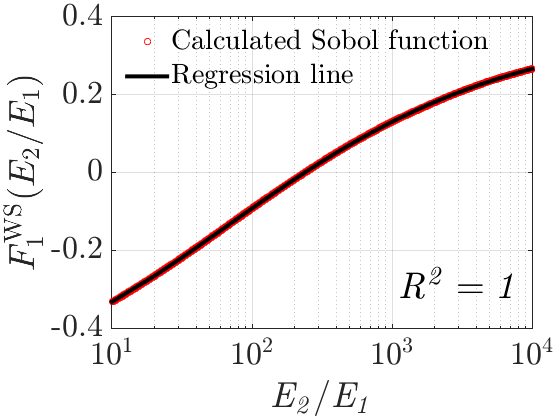}}
  \subfigure{
    \includegraphics[scale=0.5]{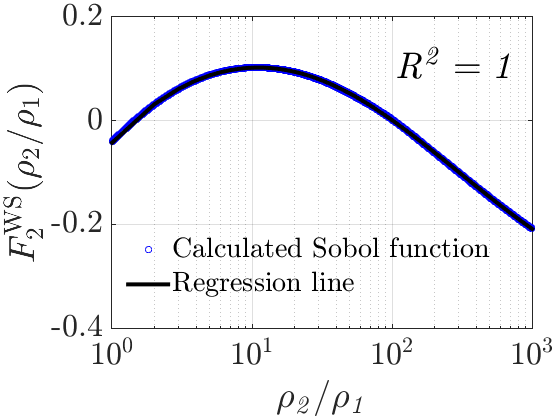}}
  \caption{Curve fitting on Sobol’ functions for the
    width of the first frequency band gap of S-Wave.
    \label{Fig15:Fit_Sobol_function_S-Wave_width}}
\end{figure}

Similarly, by fitting and summing up the dominant Sobol’ functions as shown in Figure \ref{Fig15:Fit_Sobol_function_S-Wave_width}, the reduced objective function for the width of the first frequency band gap of S-Wave can be obtained. The reduced objective function is shown in Eq.~\eqref{Eqn:Sobol_SWave_width_equation}. The fitted Sobol’ functions for Eq.~\eqref{Eqn:Sobol_SWave_width_equation} is tabulated in Table \ref{Table:Sobol_SWave_width_approx_functions}.

\begin{figure}
  \centering
  \subfigure{
    \includegraphics[scale=0.5]{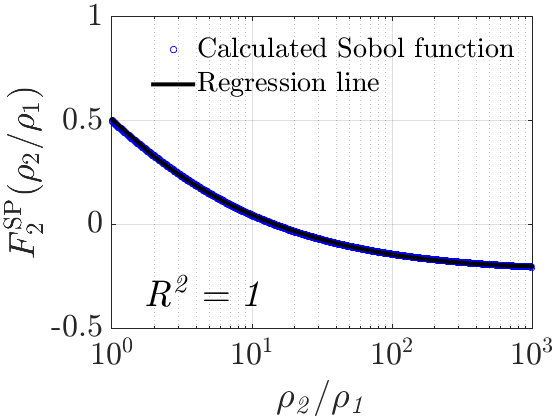}}
  \subfigure{
    \includegraphics[scale=0.5]{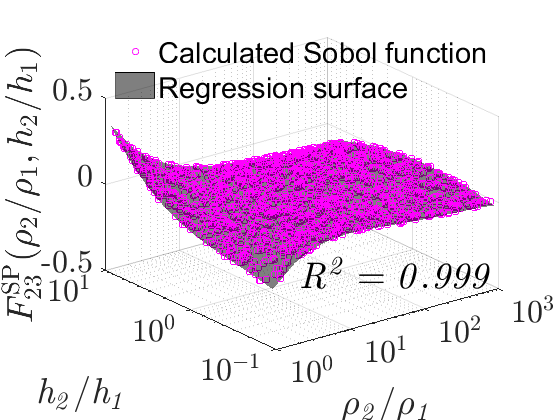}}
  \subfigure{
    \includegraphics[scale=0.5]{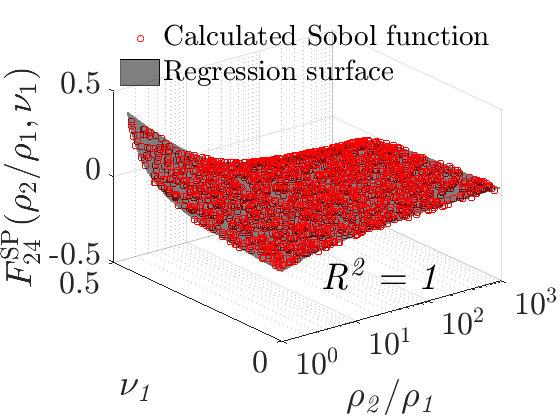}}
  \subfigure{
    \includegraphics[scale=0.5]{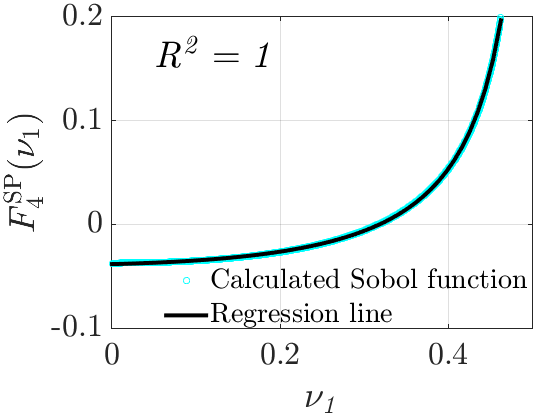}}
  \subfigure{
    \includegraphics[scale=0.5]{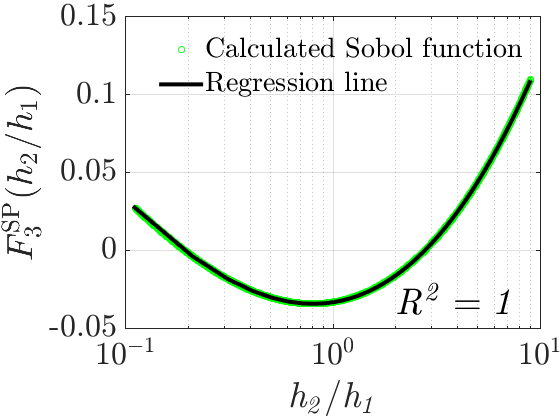}}
  \caption{Curve fitting on Sobol’ functions for the
    starting of the first frequency band gap of P-Wave.
    \label{Fig16:Fit_Sobol_function_P-Wave_start}}
\end{figure}

\begin{figure}
  \centering
  \subfigure{
    \includegraphics[scale=0.5]{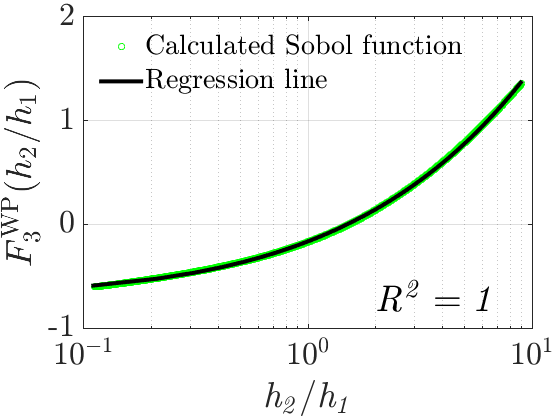}}
  \subfigure{
    \includegraphics[scale=0.5]{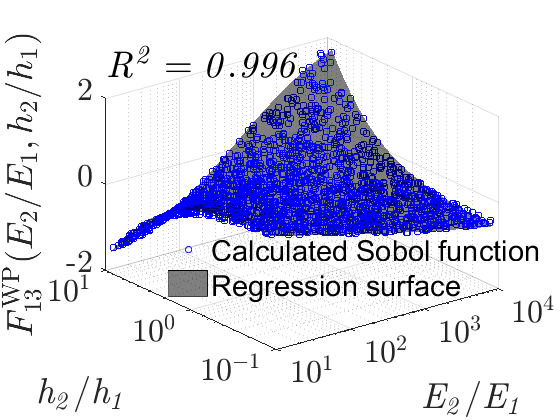}}
  \subfigure{
    \includegraphics[scale=0.5]{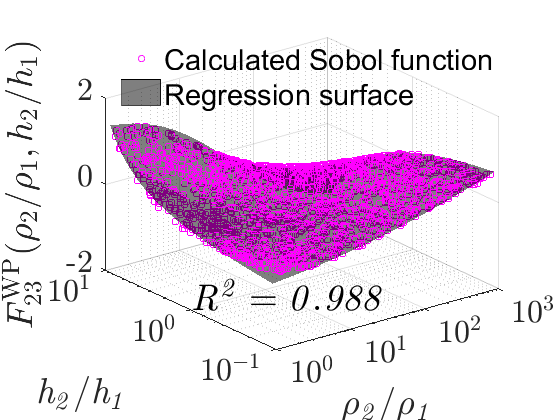}}
  \subfigure{
    \includegraphics[scale=0.5]{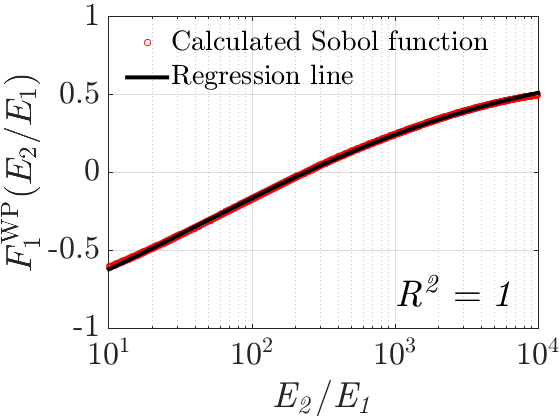}}
  \subfigure{
    \includegraphics[scale=0.5]{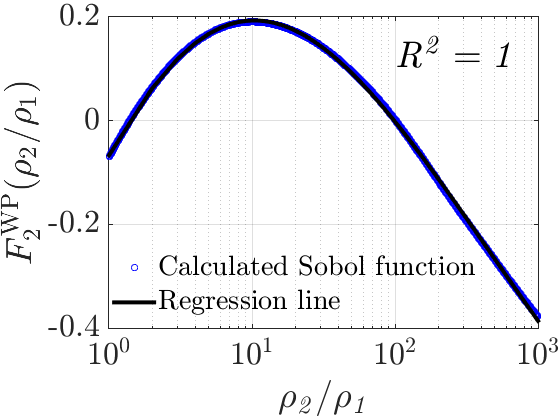}}
  \subfigure{
    \includegraphics[scale=0.5]{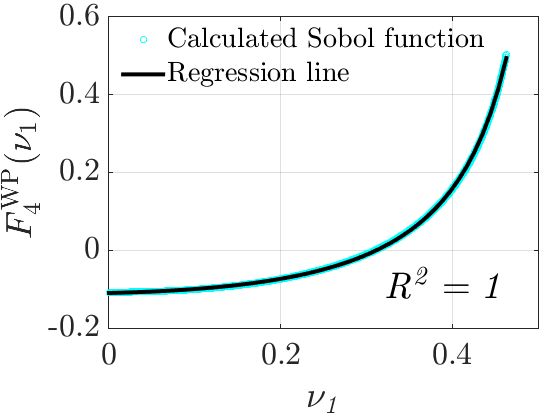}}
  \caption{Curve fitting on Sobol’ functions for the width of the first frequency band gap of P-Wave.
    \label{Fig17:Fit_Sobol_function_P-Wave_width}}
\end{figure}

The fitted functions for each of the starting and the width of the first frequency band gap subjected to P-Wave is shown in each of Figure \ref{Fig16:Fit_Sobol_function_P-Wave_start} and Figure \ref{Fig17:Fit_Sobol_function_P-Wave_width}, respectively. The reduced objective functions to predict the first frequency band gap of P-Wave are given in Eqs.~\eqref{Eqn:Sobol_PWave_start_equation} and ~\eqref{Eqn:Sobol_PWave_width_equation} with each corresponding Sobol’ functions summarized in Tables \ref{Table:Sobol_PWave_start_approx_functions} and \ref{Table:Sobol_PWave_width_approx_functions}.
Note that the functions presented in each of Eqs.~\eqref{Eqn:Sobol_SWave_start_equation} to ~\eqref{Eqn:Sobol_PWave_width_equation} are listed in decreasing order of importance.
\begin{align}
\begin{split}
  \label{Eqn:Sobol_SWave_start_equation}
  \text{Starting of }1^{\text{st}}\text{ frequency band gap (S-Wave)}
  = 0.1265 + F_2^{\text{SS}}(\log[\rho_2/\rho_1]) \\
  + F_{23}^{\text{SS}}(\log[\rho_2/\rho_1],\log[h_2/h_1])
  + F_3^{\text{SS}}(\log[h_2/h_1])
\end{split}
\end{align}
%
\begin{align}
\begin{split}
  \label{Eqn:Sobol_SWave_width_equation}
  \text{Width of }1^{\text{st}} \text{ frequency band gap (S-Wave)}
  = 0.5484 + F_3^{\text{WS}}(\log[h_2/h_1]) \\
  + F_{13}^{\text{WS}}(log[E_2/E_1],\log[h_2/h_1]) 
  + F_{23}^{\text{WS}}(\log[\rho_2/\rho_1],\log[h_2/h_1])\\
  + F_1^{\text{WS}}(\log[E_2/E_1]) + F_2^{\text{WS}}(\log[\rho_2/\rho_1])
\end{split}
\end{align}
%
\begin{align}
\begin{split}
  \label{Eqn:Sobol_PWave_start_equation}
  \text{Starting of }1^{\text{st}} \text{ frequency band gap (P-Wave)}
  = 0.2348 + F_2^{\text{SP}}(\log[\rho_2/\rho_1])\\
  + F_{23}^{\text{SP}}(\log[\rho_2/\rho_1],\log[h_2/h_1])
  + F_{24}^{\text{SP}}(\log[\rho_2/\rho_1],\upsilon_1)\\
  + F_4^{\text{SP}}(\upsilon_1) + F_3^{\text{SP}}(\log[h_2/h_1])
\end{split}
\end{align}
%
\begin{align}
\begin{split}
  \label{Eqn:Sobol_PWave_width_equation}
  \text{Width of }1^{\text{st}} \text{ frequency band gap (P-Wave)}
  = 1.0021 + F_3^{\text{WP}}(\log[h_2/h_1])\\
  +F_{13}^{\text{WP}}(\log[E_2/E_1],\log[h_2/h_1])
  +F_{23}^{\text{WP}}(\log[\rho_2/\rho_1],\log[h_2/h_1])\\
  +F_1^{\text{WP}}(\log[E_2/E_1])+F_2^{\text{WP}}(\log[\rho_2/\rho_1])+F_4^{\text{WP}}(\upsilon_1)
\end{split}
\end{align}

\begin{table}
  \caption{Regression equation for Sobol’
    functions for the starting of the
    first frequency band gap of S-Wave
    \label{Table:Sobol_SWave_start_approx_functions}}
  \begin{tabular}{|c|} \hline
    \\[-10pt] 
    $F_2^{\text{SS}}(X)=1239.088 e^{-0.4557X}
    - 1238.816 e^{-0.4555X}$ \\ \\[-12pt] \hline
    \\[-10pt]
    $F_{23}^{\text{SS}}(X,Y) = -0.02746
    + 0.02426 X + 0.1143 Y$\\
    $-0.001248 X^2 - 0.2258 XY +0.09419 Y^2$ \\
    $- 0.001116 X^3 + 0.1204 X^2 Y - 0.09103 X Y^2$ \\
    $+ 0.0001409 X^4 - 0.02031 X^3 Y 
    + 0.0142 X^2 Y^2$ \\ \\[-12pt] \hline
    \\[-10pt]
    $F_3^{\text{SS}}(Y) = -0.01822 + 0.0114 Y 
    + 0.06029 Y^2 + 0.01339Y^3$
    \\[-12pt] \\
    \hline 
  \end{tabular}
\end{table}

\begin{table}
  \caption{Regression equation for Sobol’ functions for the width of the first frequency band gap of S-Wave
   \label{Table:Sobol_SWave_width_approx_functions}}
  \begin{tabular}{|c|} \hline
    \\[-10pt]
    $F_3^{\text{WS}}(Z)=\dfrac{-0.4961+2.7538Z}{5.2843-3.5212Z+Z^2}$ \\  \\[-12pt] \hline
    \\[-10pt]
    $F_{13}^{\text{WS}}(X,Z)=\dfrac{0.2233-0.08928X-0.785Z+0.3128XZ}{1-0.2172X-0.7975Z+0.04177X^2+0.3198Z^2}$ \\ \\[-12pt] \hline
    \\[-10pt]
    $F_{23}^{\text{WS}}(Y,Z)=\dfrac{-0.1193+0.0735Y+0.4151Z-0.2566YZ}{1-0.2974Y-0.8122Z+0.08275YZ+0.0607Y^2+0.1766Z^2}$ \\ \\[-12pt] \hline
    \\[-10pt]
    $F_1^{\text{WS}}(X)=\dfrac{-5.1888+1.7591X+0.1783X^2}{11.1718-2.3914X+X^2}$ \\ \\[-12pt] \hline
    \\[-10pt]
    $F_2^{\text{WS}}(Y)=\dfrac{-0.8474+5.817Y-2.6891Y^2}{19.9635+3.7597Y-2.4232Y^2+Y^3}$ \\[-12pt] \\
    \hline 
  \end{tabular}
\end{table}

\begin{table}
  \caption{Regression equation for Sobol’ functions for the starting of the first frequency band gap of P-Wave
   \label{Table:Sobol_PWave_start_approx_functions}}
  \begin{tabular}{|c|} \hline
    \\[-10pt]
    $F_2^{\text{SP}}(X)=297.7911e^{(-0.4565X)}-297.2854e^{(-0.4551X)}$ \\ \\[-12pt] \hline
    \\[-10pt]
    $F_{23}^{\text{SP}}(X,Y)=-0.04856+0.03971X+0.1815Y+0.00122X^2-0.5028XY$\\
				     $+0.1629Y^2-0.003424X^3+0.4332X^2Y-0.1377XY^2$\\
				     $+0.1008Y^3+0.00077X^4-0.1538X^3Y+0.004778(XY)^2$\\
				     $-0.1301XY^3-0.000102X^5+0.01921X^4Y$\\
				     $+0.004352X^3Y^2+0.03148X^2Y^3$ \\ \\[-12pt] \hline
    \\[-10pt]
    $F_{24}^{\text{SP}}(X,Z)=\dfrac{-0.0969+0.08431X+0.3088Z-0.2676XZ}{1+0.847X-1.9226Z-1.7922XZ+0.02933X^2}$\\ \\[-12pt] \hline
    \\[-10pt]
    $F_4^{\text{SP}}(Z)=\dfrac{-209.3429+507.5532Z+471.1006Z^2}{5462.1354-10408.7099Z+Z^2}$\\  \\[-12pt]\hline
    \\[-10pt]
    $F_3^{\text{SP}}(Y)=-0.03341+0.02097Y+0.1116Y^2+0.02439Y^3$ \\[-12pt] \\
    \hline 
  \end{tabular}
\end{table}

\begin{table}
  \caption{Regression equation for Sobol’ functions for the width of the first frequency band gap of P-Wave
   \label{Table:Sobol_PWave_width_approx_functions}}
  \begin{tabular}{|c|} \hline
    \\[-10pt]
    $F_3^{\text{WP}}(Y)=1.1042e^{(1.0456Y)}-1.2681e^{(0.2525Y)}$\\ \\[-12pt] \hline
    \\[-10pt]
    $F_{13}^{\text{WP}}(W,Y)=\dfrac{0.4458-0.1779W-1.5468Y+0.6325WY-0.06565Y^2}{1-0.2054W-0.684Y-0.067WY+0.04826W^2+0.3312Y^2}$ \\ \\[-12pt] \hline
    \\[-10pt]
    $F_{23}^{\text{WP}}(X,Y)=\dfrac{-0.1734+0.1082X+0.6911Y-0.4181XY-0.04576Y^2}{1-0.2716X-0.7256Y+0.1226XY+0.04735X^2}$ \\ \\[-12pt] \hline
    \\[-10pt]
    $F_1^{\text{WP}}(W)=\dfrac{-26139.7298+11020.821W}{26761.2895-3942.3173W+1499W^2+W^3}$ \\ \\[-12pt] \hline
    \\[-10pt]
    $F_2^{\text{WP}}(X)=\dfrac{-1383.2386+10241.098X-4787.7159X^2}{20757.67-2034.7302X+2324.8446X^2+X^3}$\\ \\[-12pt] \hline
    \\[-10pt]
    $F_4^{\text{WP}}(Z)=\dfrac{-296.2129+679.2197Z+871.5131Z^2}{2697.0582-4930.097Z+Z^2}$\\[-12pt] \\
    \hline 
  \end{tabular}
\end{table}

The accuracy of the reduced objective functions depend on both the truncation of the Sobol’s function series and the regression chosen to represent each function. To quantify the approximation error of the reduced objective functions, the scaled $L_2$ error ($\delta$) is used.

\begin{align}
  \label{Eqn:Sobol_L2_error}
  \delta=\dfrac{1}{D}\int[F_{\text{response}}(\textbf{x})-F_{\text{predict}}(\textbf{x})]dx
\end{align}

\begin{figure}
  \centering
  \includegraphics[scale=0.6]{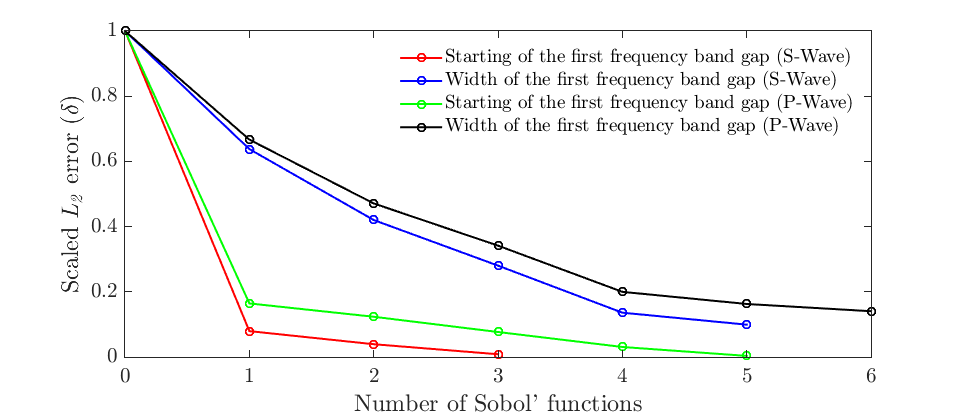}
  \caption{Scaled $L_{2}$ errors of the reduced objective functions.
    \label{Fig18:L2_errorl}}
\end{figure}

The scaled $L_2$ error specifies how far the predicted or reduced objective function from the real response function. If the crudest approximation $F_{\mathrm{predict}}=F_0$, then $\delta=1$. Hence, the best approximations are the ones with $\delta\ll1$. The scaled $L_2$ error should be interpreted as the combination of truncation error and regression error. Figure \ref{Fig18:L2_errorl} shows the evolution of the $L_2$ error with respect to the number of Sobol’ functions included in the reduced objective functions. The $L_2$ error was evaluated using 2000 input samples. The sequence of the included Sobol’ functions is as presented in Eqs.~\eqref{Eqn:Sobol_SWave_start_equation} to ~\eqref{Eqn:Sobol_PWave_width_equation}. Since the starting of the frequency band gap for both S-Wave and P-wave are highly depend on their respected first Sobol’ functions, huge drop in the $L_2$ error is observed after the inclusion of the first functions (see red and green curves in Figure \ref{Fig18:L2_errorl}). Inclusion of additional Sobol’ functions can reduced the error but they are not much significant. As for the width of the first frequency band gaps of S-Wave and P-Wave, the error can be seen gradually decreased by inclusion of more Sobol’ functions. For the case of S-Wave, the inclusion of the fifth Sobol’ function only slightly reduce the error, as shown by the blue curve. Similarly for the case of P-Wave, the inclusion of the sixth Sobol’ function also slightly reduce the error.

The reduced objective functions, therefore, can predict the starting frequency of the first frequency band gaps of S-Wave and P-Wave very accurately, as the scaled $L_2$ error pointing very small values of 0.0086 and 0.0036, respectively. For the case of width of the first frequency band gap, the scaled $L_2$ error show small values of 0.0994 and 0.1406, respectively, for S-Wave and P-Wave, which is very acceptable from engineering point of view. Depending on the level of accuracy, one may include more Sobol’ functions in the reduced objective functions to further diminish the truncation error.

The reduced objective functions can be used as simplified design equations to easily obtain the first frequency band gaps for S-Wave and P-Wave. Note that the design equations provide the frequencies in non-dimensional unit. To obtain the dimension in Hz, the calculated non-dimensional frequencies need to be divided by the reference variable $T_*$.

\section{CLOSURE}
\label{Sec:S5_Sobol_CR}
This paper presented a global sensitivity
analysis based on variance decomposition
(i.e., Sobol’ sensitivity analysis) to
characterize the most influential parameters
in designing frequency band gaps of periodic
materials. We
focused our attention on the first frequency band gap of two-layer 1D phononic crystals subjected to each of S-Wave and P-Wave as the objective functions. The input parameters are selected as the ratio of material and geometric properties of the two layers that compose the unit cell.
The Sobol' analysis revealed the following: 
\begin{enumerate}[(i)]
\item For the lower bound frequency or
  the starting of frequency band gap
  subjected to S-Wave, it is observed
  that the density ratio is the predominant
  parameter followed by the interaction of
  density ratio and thickness ratio parameters
  and the single parameter of thickness ratio.
\item For the starting frequency band gap
  under P-Wave, the affecting parameters
  are the same as those when subjected to S-Wave with addition of the interaction of density ratio and first Poisson’s ratio parameters and the single parameter of first Poisson’s ratio.
\item For the width of frequency band gaps,
  more parameters and the interaction of
  these parameters had significant affect
  on the objective functions. In the case of S-Wave, the dominant parameters are the thickness ratio, the interaction of Young’s modulus ratio and thickness ratio, the interaction of density ratio and thickness ratio, the Young’s modulus ratio and the density ratio. In the case of P-Wave, the dominant parameters are the same as those of S-Wave with addition of the first Poison’s ratio parameter. 
\end{enumerate}

The ability of Sobol’ sensitivity analysis to assess the interaction between parameters and how they affect the objective functions, can provide the insight and better understanding into uncertainty in the model, which set this method apart from standard sensitivity analysis.
Finally, guided by the Sobol’ indices, simplified design equations with reduced number of input were proposed. The design equations can quickly estimate the first frequency band gap of two-layer 1D phononic crystals subjected to S-Wave and P-Wave without the need to solve the wave equation. The error analysis using 2000 input samples show that the proposed design equation can predict the first frequency band gap with good accuracy. In fact, one may modify the proposed equation by adding more or reducing the Sobol’ function terms depending on the desired level of accuracy.

\section*{Acknowledgments}
We thank the financial support from the U.S. Department of Energy NEUP 
program (Project No. CFA-14-6446). The opinions 
expressed in this study are those of the authors and 
do not necessarily reflect the views of the sponsor.
The authors also extend their gratitude to Kuo-Chun Chang 
(National Taiwan University), Yu Tang (Argonne National Laboratory) 
and Robert Kassawara (Electric Power Research Institute) for their 
contribution to the experiment component of the NEUP project; which 
complements the theoretical study presented in this paper.

\bibliographystyle{plainnat}
\bibliography{References}
\newpage

\end{document}